\documentclass[twocolumn]{aastex631}

\usepackage{savesym}
\savesymbol{tablenum}
\usepackage{siunitx}
\restoresymbol{SIX}{tablenum}
\usepackage{soul}

\newcommand{\hii}{\mbox{\ion{H}{2}}}
\newcommand{\oii}{\mbox{[\ion{O}{2}]}}
\newcommand{\oiii}{\mbox{[\ion{O}{3}]}}

\newcommand{\hei}{\mbox{\ion{He}{1}}}
\newcommand{\nii}{\mbox{[\ion{N}{2}]}}
\newcommand{\sii}{\mbox{[\ion{S}{2}]}}
\newcommand{\hb}{\mbox{H$\beta$}}
\newcommand{\ha}{\mbox{H$\alpha$}}
\newcommand{\lya}{\mbox{Ly$\alpha$}}

\newcommand\msun{\mbox{\si{M_\odot}}}

\newcommand\smpy{\mbox{\si{M_\odot.yr^{-1}}}}
\newcommand\mstar{\mbox{$M_\mathrm{star}$}}

\renewcommand{\micron}{\si{\micro\meter}}
\newcommand{\zsun}{\si{Z_\odot}}

\defcitealias{sun22}{\ion{Paper}{1}}

\newcommand{\RomanNumeralCaps}[1]{\uppercase\expandafter{\romannumeral#1}}

\newcommand{\textred}[1]{{#1}}

\received{September 7, 2022}
\revised{May 10, 2023}
\accepted{May 11, 2023}

\shorttitle{First Sample of \ha$+$\oiii\ Emitters at $z>6$ through JWST/NIRCam Grism}
\shortauthors{Sun et al.}

\begin{document}

\title{First Sample of \ha$+$\oiii\,$\lambda$5007 Line Emitters at $\mathbf{z>6}$ Through JWST/NIRCam Slitless Spectroscopy: Physical Properties and Line Luminosity Functions}

\author[0000-0002-4622-6617]{Fengwu Sun}
\affiliation{Steward Observatory, University of Arizona, 933 N. Cherry Avenue, Tucson, AZ 85721, USA}

\author[0000-0003-1344-9475]{Eiichi Egami}
\affiliation{Steward Observatory, University of Arizona, 933 N. Cherry Avenue, Tucson, AZ 85721, USA}

\author[0000-0003-3382-5941]{Nor Pirzkal}
\affiliation{ESA/AURA STScI, 3700 San Martin Dr., Baltimore, MD 21218, USA}

\author[0000-0002-7893-6170]{Marcia Rieke}
\affiliation{Steward Observatory, University of Arizona, 933 N. Cherry Avenue, Tucson, AZ 85721, USA}

\author[0000-0002-4735-8224]{Stefi Baum}
\affiliation{Dept of Physics \& Astronomy, University of Manitoba, Winnipeg MB R3T 2N2 Canada}

\author[0000-0003-4850-9589]{Martha Boyer}
\affiliation{Space Telescope Science Institute, 3700 San Martin Drive, Baltimore, MD 21218, USA}

\author[0000-0003-4109-304X]{Kristan Boyett}
\affiliation{School of Physics, University of Melbourne, Parkville 3010, VIC, Australia}
\affiliation{ARC Centre of Excellence for All Sky Astrophysics in 3 Dimensions (ASTRO 3D), Australia}

\author[0000-0002-8651-9879]{Andrew J.\ Bunker}
\affiliation{University of Oxford, Department of Physics, Denys Wilkinson Building, Keble Road, Oxford OX13RH, UK}

\author[0000-0002-0450-7306]{Alex J.\ Cameron}
\affiliation{Sub-department of Astrophysics, University of Oxford, Keble Road, Oxford OX1 3RH, United Kingdom}

% \author[0000-0001-6464-3257]{Matteo Correnti}
% \affiliation{Space Telescope Science Institute, 3700 San Martin Drive, Baltimore, MD 21218, USA}

\author[0000-0002-2678-2560]{Mirko Curti}
\affiliation{Kavli Institute for Cosmology, University of Cambridge, Madingley Road, Cambridge, CB3 0HA, UK}
\affiliation{Cavendish Laboratory - Astrophysics Group, University of Cambridge, 19 JJ Thompson Avenue, Cambridge, CB3 0HE, UK}

\author[0000-0002-2929-3121]{Daniel J.\ Eisenstein}
\affiliation{Center for Astrophysics $|$ Harvard \& Smithsonian, 60 Garden St., Cambridge MA 02138 USA}

\author[0000-0002-5581-2896]{Mario Gennaro}
\affiliation{Space Telescope Science Institute, 3700 San Martin Drive, Baltimore, MD 21218, USA}
\affiliation{Department of Physics and Astronomy, Johns Hopkins University, 3400 North Charles Street, Baltimore, MD 21218, USA}

% \author[0000-0001-8627-0404]{Julien Girard}
% \affiliation{Space Telescope Science Institute, 3700 San Martin Drive, Baltimore, MD 21218, USA}

\author[0000-0002-8963-8056]{Thomas P. Greene}
\affiliation{Space Science and Astrobiology Division, NASA Ames Research Center, MS 245-6, Moffett Field, CA 94035 USA}

\author[0000-0003-3577-3540]{Daniel Jaffe}
\affiliation{Department of Astronomy, University of Texas at Austin, 2515 Speedway Blvd Stop C1400, Austin, TX 78712, USA}

\author{Doug Kelly}
\affiliation{Steward Observatory, University of Arizona, 933 N. Cherry Avenue, Tucson, AZ 85721, USA}

\author[0000-0002-6610-2048]{Anton M. Koekemoer}
\affiliation{Space Telescope Science Institute, 3700 San Martin Drive, Baltimore, MD 21218, USA}

\author[0000-0002-5320-2568]{Nimisha Kumari}
\affiliation{AURA for the European Space Agency (ESA), Space Telescope Science Institute, 3700 San Martin Drive, Baltimore, MD 21218, USA}

% \author[0000-0002-0834-6140]{Jarron Leisenring}
% \affiliation{Steward Observatory, University of Arizona, 933 N. Cherry Avenue, Tucson, AZ 85721, USA}

\author[0000-0002-4985-3819]{Roberto Maiolino}
\affiliation{Kavli Institute for Cosmology, University of Cambridge, Madingley Road, Cambridge, CB3 0HA, UK}
\affiliation{Cavendish Laboratory - Astrophysics Group, University of Cambridge, 19 JJ Thompson Avenue, Cambridge, CB3 0HE, UK}
\affiliation{Department of Physics and Astronomy, University College London, Gower Street, London WC1E 6BT, UK}

\author[0000-0003-0695-4414]{Michael Maseda}
\affiliation{Department of Astronomy, University of Wisconsin-Madison, 475 N. Charter Street, Madison, WI 53706, USA}

% \author{Karl Misselt}
% \affiliation{Steward Observatory, University of Arizona, 933 N. Cherry Avenue, Tucson, AZ 85721, USA}

% \author[0000-0002-6500-3574]{Nikolay Nikolov}
% \affiliation{Space Telescope Science Institute, 3700 San Martin Drive, Baltimore, MD 21218, USA}

\author[0000-0002-0362-5941]{Michele Perna}
\affiliation{Centro de Astrobiolog\'ia (CAB), CSIC-INTA, Ctra. de Ajalvir Km. 4, 28850 Torrej\'on de Ardoz, Madrid, Spain}

\author[0000-0002-4410-5387]{Armin Rest}
\affiliation{Space Telescope Science Institute, 3700 San Martin Drive, Baltimore, MD 21218, USA}
\affiliation{Department of Physics and Astronomy, Johns Hopkins University, 3400 North Charles Street, Baltimore, MD 21218, USA}

\author[0000-0002-4271-0364]{Brant E. Robertson}
\affiliation{Department of Astronomy and Astrophysics, University of California, Santa Cruz, 1156 High Street, Santa Cruz, CA 95064, USA}

% \author[0000-0002-6730-5410]{Thomas L. Roellig}
% \affiliation{NASA Ames Research Center, MS 245-6, Moffett Field, CA 94035, USA}

\author[0000-0001-8291-6490]{Everett Schlawin}
\affiliation{Steward Observatory, University of Arizona, 933 N. Cherry Avenue, Tucson, AZ 85721, USA}

\author[0000-0001-8034-7802]{Renske Smit}
\affiliation{Astrophysics Research Institute, Liverpool John Moores University, 146 Brownlow Hill, Liverpool L3 5RF, UK}

\author[0000-0003-2434-5225]{John Stansberry}
\affiliation{Space Telescope Science Institute, 3700 San Martin Drive, Baltimore, MD 21218, USA}

\author[0000-0003-3759-8707]{Ben Sunnquist}
\affiliation{Space Telescope Science Institute, 3700 San Martin Drive, Baltimore, MD 21218, USA}

\author[0000-0002-8224-4505]{Sandro Tacchella}
\affiliation{Kavli Institute for Cosmology, University of Cambridge, Madingley Road, Cambridge, CB3 0HA, UK}
\affiliation{Cavendish Laboratory - Astrophysics Group, University of Cambridge, 19 JJ Thompson Avenue, Cambridge, CB3 0HE, UK}

\author[0000-0003-2919-7495]{Christina C. Williams}
\affiliation{NSF's National Optical-Infrared Astronomy Research Laboratory, 950 North Cherry Avenue, Tucson, AZ 85719, USA}
\affiliation{Steward Observatory, University of Arizona, 933 N. Cherry Avenue, Tucson, AZ 85721, USA}

\author[0000-0001-9262-9997]{Christopher N. A. Willmer}
\affiliation{Steward Observatory, University of Arizona, 933 N. Cherry Avenue, Tucson, AZ 85721, USA}

% \author{other JADES \& JWST/NIRCam Commissioning Team Members}

% \collaboration{18}{(Members of the JWST/NIRCam Commissioning Team)}

\correspondingauthor{Fengwu Sun}
\email{fengwusun@arizona.edu}



\begin{abstract}
We present a sample of four emission-line galaxies at $z=6.11-6.35$ that were serendipitously discovered using the commissioning data for the JWST/NIRCam wide-field slitless spectroscopy (WFSS) mode.
One of them (at $z$\,$=$\,6.11) has been reported previously, while the others are new discoveries.
These sources are selected by the secure detections of both \oiii\,$\lambda$5007 and \ha\ lines with other fainter lines tentatively detected in some cases (e.g., \oii\,$\lambda$3727, \oiii\,$\lambda$4959).
In the \oiii/\hb--\nii/\ha\ Baldwin-Phillips-Terlevich diagram, these galaxies occupy the same parameter space as that of $z\sim2$ star-forming galaxies, indicating that they have been enriched rapidly to subsolar metallicities (\textred{$\sim0.4\,\mathrm{Z}_{\odot}$}), similar to galaxies with comparable stellar masses at much lower redshifts.
The detection of strong \ha\ lines suggests a higher ionizing photon production efficiency within galaxies in the early Universe.
We find brightening of the \oiii\,$\lambda$5007 line luminosity function (LF) from $z=3$ to 6, and \textred{weak or no} redshift evolution of the \ha\ line LF from $z=2$ to 6.  
Both LFs are under-predicted at $z\sim6$ by a factor of $\sim$10 in certain cosmological simulations.
This further indicates a global \lya\ photon escape fraction of \textred{7--10\%} at $z\sim6$, \textred{slightly} lower than previous estimates through the comparison of the UV-derived star-formation rate density and \lya\ luminosity density.
Our sample recovers \textred{$66_{-44}^{+128}$}\%\ of $z = 6.0-6.6$ galaxies in the survey volume with stellar masses greater than $5\times10^{8}\,$\msun, suggesting the ubiquity of strong \ha\ and \oiii\ line emitters in the Epoch of Reionization, which will be further uncovered in the era of JWST.
\end{abstract}



\section{Introduction} \label{sec:01_intro}

With eighteen images of the same star from each segment of the primary mirror finally aligned together, the long-awaited James Webb Space Telescope (JWST; \textred{\citealt{gardner23}}) immediately started to unfold the secrets from the distant Universe.
Thanks to its unprecedented sensitivity and spectroscopic capability in the near/mid-infrared (NIR/MIR) wavelengths, for the first time the rest-frame optical nebular emission lines (e.g., \oii\,$\lambda\lambda$3726, 3729, \hb, \oiii\,$\lambda\lambda$4959, 5007 and \ha) of normal star-forming galaxies can be directly detected and resolved in the Epoch of Reionization (EoR: $z$\,$\gtrsim$\,6; see a recent review by \citealt{robertson22}).  
The JWST/NIRSpec Early Release Observations (EROs) of the SMACS0723 lensing-cluster field \citep{pontoppidan22} have immediately produced high-quality spectra for a handful of $z\sim8$ galaxies with low stellar masses ($M_\mathrm{star} \lesssim 10^8$\,\msun).
This allows numerous studies, including direct gas-phase metallicity measurements (see \citealt{ac22}, \citealt{brinchmann22}, \citealt{carnall22}, \citealt{curti22}, \citealt{katz22}, \citealt{rhoads22}, \citealt{schaerer22}, \citealt{tacchella22}, \citealt{taylor22}, \citealt{trump22}).  
These ERO results have clearly demonstrated the power of JWST's spectroscopic observations, initially with NIRSpec, promising many exciting discoveries to be made over the coming years.

In addition to NIRSpec, NIRCam's wide-field slitless spectroscopy (WFSS; \citealt{greene17}, \textred{\citealt{rieke23}}) mode offers a uniquely powerful capability, allowing us to conduct blind (i.e., unbiased) surveys of EoR galaxies with strong line emissions in the rest-frame optical.  
Studies in recent years have shown that there exists a substantial population of star-forming galaxies at $z\gtrsim6$ whose rest-frame emission lines may be strong enough to distort Spitzer/IRAC [3.6]--[4.5] \micron\ broad-band colors \citep[e.g.,][]{egami05, schaerer09,stark13,labbe13,smit14, smit15, rb16, harikane18, endsley21b, endsley21a}. 
% This means that 
Although NIRCam/Grism spectroscopy is not as sensitive as that of NIRSpec because of a higher background, the NIRCam WFSS mode has the potential to sample a substantial number of line-emitting galaxies in the EoR. 
Indeed, this has become a real possibility with the serendipitous discovery of a line emitter at $z=6.11$ with bright \oiii\,$\lambda$5007 and \ha\ emission lines in the shallow ($\sim$\,20\,min\ integration) NIRCam WFSS commissioning data (\citealt{sun22}; hereafter \citetalias{sun22}), confirming the ubiquity of galaxies with strong rest-frame optical emission in the EoR.

Scientifically, a unique power of NIRCam WFSS data is its ability to directly measure or constrain the luminosity functions (LFs) of strong rest-frame optical lines at $z\gtrsim6$.
However, this will be difficult with NIRSpec multi-object spectroscopy (MOS) observations because of (\romannumeral1) complex target selection functions inherent in various criteria based on continuum colors and (\romannumeral2) wavelength-dependent slit losses.
This is also impossible with NIRISS WFSS observations because of a shorter wavelength coverage (0.8--2.2\,\micron, in contrast to 2.4--5.0\,\micron\ of NIRCam WFSS).  
The emergence of strong line emitters inferred from the IRAC observations mentioned above suggests a significant brightening of the corresponding line LF toward high redshift.  
Indeed, a recent estimate of the $z\sim8$ \oiii$+$\hb\ LF by \citet{debarros19} does show such a trend.  
This is in great contrast to the rapid dimming seen with the Lyman\,$\alpha$ (\lya) LF \citep[e.g.,][]{konno18} and that of UV continuum \citep[e.g.,][]{bouwens16b,bouwens21a}.  
While the former reflects absorption because of increasingly neutral IGM at high redshift, the combination of increasing \oiii$+$\hb\ and decreasing UV luminosities would imply a systematic increase of the hydrogen ionizing production efficiency ($\xi_\mathrm{ion}$) in these galaxies (i.e., more hydrogen atoms are ionized for a given UV luminosity), which has been directly observed in strong \oiii\ emitters at $z=1.3-2.4$ \citep[e.g.,][]{tang19}.
NIRCam WFSS surveys targeting EoR galaxies will enable accurate and robust determination of line LFs in the rest-frame optical, which will allow us to investigate these issues directly.

%% This work

In this work, we present a sample of four  \oiii\,$\lambda$5007 and \ha\ line emitters at $z > 6$ that were discovered serendipitously with the JWST/NIRCam WFSS mode.
These galaxies were discovered in the field around the flux-calibration star P330-E (GSC\,02581-02323), which was observed during the commissioning phase of this observing mode. 
Among them, the lowest-redshift source at $z=6.11$ has been reported in \citetalias{sun22}, and the other three sources at $z=6.15-6.35$ are reported for the first time.
The detections of both \oiii\ and \ha\ emission lines for all sources provide secure spectroscopic redshift determination, and also enable the determination of their physical properties and emission-line LF.

This paper is arranged as follows. 
In Section~\ref{sec:02_obs}, we describe the JWST/NIRCam observations and corresponding data reduction techniques.
The spectroscopic and photometric measurements are presented in Section~\ref{sec:03_res}.
In Section~\ref{sec:04_disc}, we discuss the physical properties of these emission-line galaxies including the line strengths, metallicities, ionizing photon production efficiency, and the \oiii/\hb-\nii/\ha\ Baldwin--Phillips--Terlevich (BPT; \citealt{bpt1981}) diagram of galaxies at $z>6$.
In Section~\ref{sec:05_vol}, we discuss the volume density of these galaxies in the EoR, presenting the first direct measurements of the \oiii\,$\lambda$5007 and \ha\ line LFs at $z>6$.
The conclusions can be found in Section~\ref{sec:06_con}.
Throughout this paper, we assume a flat $\Lambda$CDM cosmology with $H_{0} = 70$\,\si{km.s^{-1}.Mpc^{-1}} and $\Omega_m = 0.3$, and a \citet{chabrier03} initial mass function (IMF).
The AB magnitude system \citep{oke83} is used.

\section{Observation and Data Reduction} \label{sec:02_obs}

\begin{figure*}[!t]
\centering
\includegraphics[width=\linewidth]{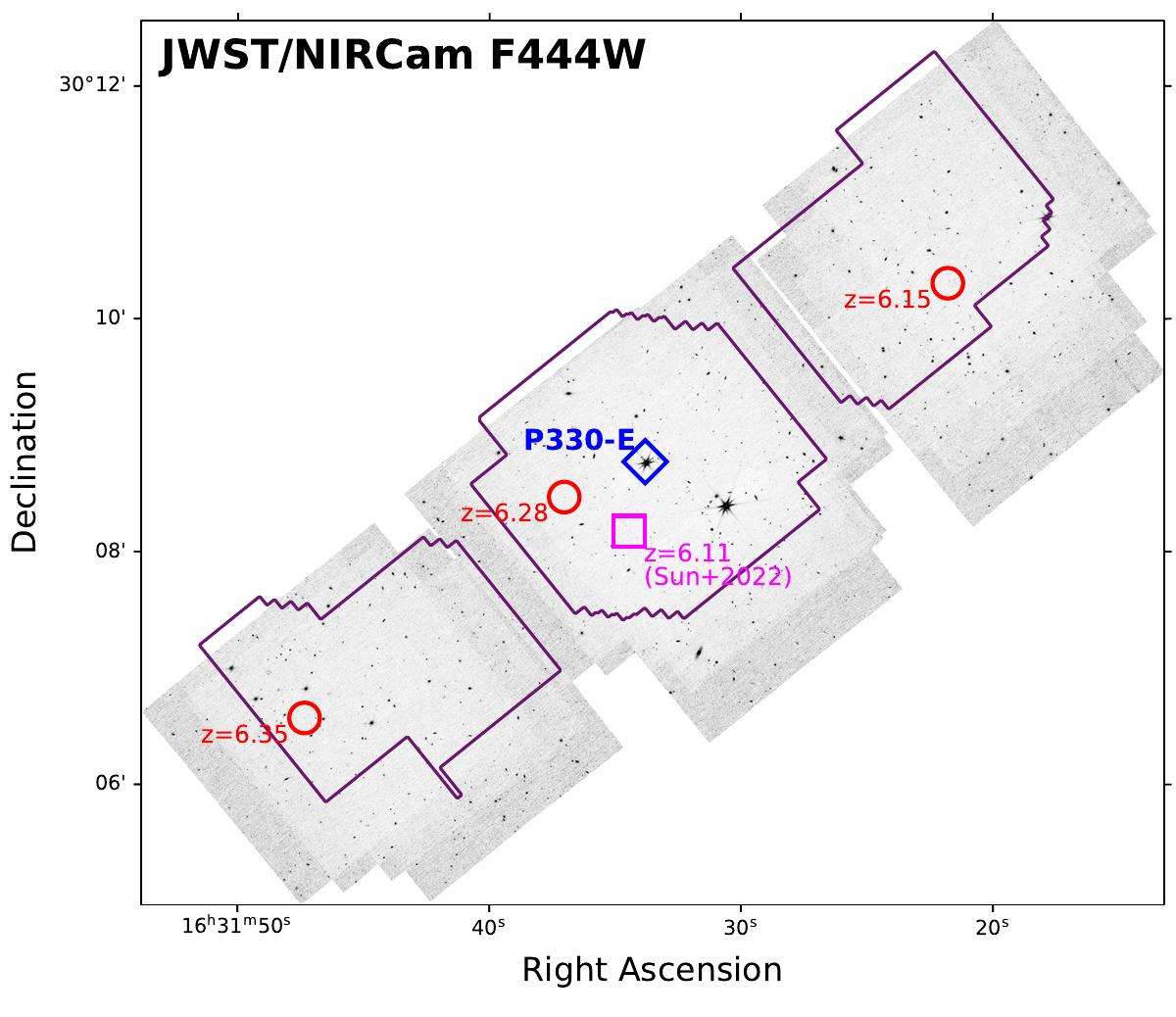}
\caption{
JWST/NIRCam F444W map of the P330-E field.
The open blue diamond denotes the primary target (P330-E as the flux calibrator) of the obtained observations.
The $z=6.11$ \oiii\ and \ha\ line emitter discovered by \citet[\citetalias{sun22}]{sun22} is shown in the open magenta square, and the three newly discovered \oiii\ and \ha \ line emitters (this work, $z=6.15-6.35$) are shown in open red circles.
The purple contours denote the regions in which sources can potentially yield both \oiii\ and \ha\ line detections with this survey at $z=6.2$ (see Appendix~\ref{apd:01}).
}
\label{fig:01_map}
\end{figure*}

The JWST/NIRCam long-wavelength (LW; 2.4--5.0\,\micron) grism characterization observations were conducted through Program \#1076 (PI: Pirzkal) during the commissioning phase of the instrument \textred{\citep{rieke23,rigby22}}.
Both grism spectroscopic and direct-imaging observations were performed with the flux-calibration star P330-E (a solar analog) and the wavelength-calibration star IRAS\,05248-7007 (a post-Asymptotic-Giant-Branch star in the Large Magellanic Cloud).
The obtained data were also described in \citetalias{sun22}.

\subsection{Direct Imaging}
\label{ss:02a_img}

% NIRCam has four grisms: each of the two modules (A and B) has two grisms that disperse light parallel to detector rows (Grism R) or columns (Grism C), respectively.
% For the flux and wavelength calibration, P330-E (a G2V star) and IRAS\,05248-7007 (a post-Asymptotic-Giant-Branch star in the Large Magellanic Cloud) were observed, respectively.

The imaging data of the P330-E field were taken either simultaneously with the grism exposures (short-wavelength, SW filter: F212N) or after the grism observations through direct and out-of-field imaging (SW filter: F212N; LW filter: F250M, F322W2, F335M and F444W).
Among them, the shallow F250M and F335M data were taken with the observations \#103/104 as part of the early spectral calibration on UT 2022 April 5.  
The total integration times were 6.4 and 4.3 min (three integrations each), respectively.
The F322W2 and F444W imaging data were taken through the observations \#105-108 and \#109-112, respectively.
The total integration time was 4\,min (eleven integrations each) with either of the two filters.

For the shallow F212N data, we only used 31 integrations associated with the LW grism exposures because the integrations were longer (96\,s each).
Shorter F212N integrations (21\,s each, \textred{8\,min in total}) taken with out-of-field imaging were not included because they caused problems for the \textsc{tweakreg} step in \textsc{jwst}\footnote{\href{https://github.com/spacetelescope/jwst}{https://github.com/spacetelescope/jwst}} Stage-3 mosaicking pipeline in source identification and world-coordinate-system (WCS) registration.
The final mosaicked F212N image has a total integration time of 50\,min.

The direct-imaging data were reduced and mosaicked using a modified Stage-1/2/3 \textsc{jwst} pipeline \textred{1.8.2} and \textsc{crds} calibration reference file context ``\textred{\texttt{jwst\_1041.pmap}''.
This version has included the Cycle-1 NIRCam photometric zero-points (see \citealt{boyer22} and \citealt{rigby22}).}
The so-called 1/f noise (see \citealt{schlawin20}) was modeled and removed using the code \textsc{tshirt/roeba}\footnote{\href{https://github.com/eas342/tshirt}{https://github.com/eas342/tshirt}} for stage-2 products (i.e., the ``\texttt{\_cal.fits}'' files). 
The ``snowball'' artifacts from cosmic rays were identified and masked through their large-area jump-detection information ($\geq$60 native pixels) on the data-quality map of each individual integration.
The final image products were resampled to a native pixel size of 0\farcs0312 (SW) and 0\farcs0629 (LW) with \texttt{pixfrac}\,$=$\,0.8,
and the WCS of the images were registered with the \textsc{Gaia} DR2 catalog \citep{gaiadr2}.  
More specifically, we registered the F444W mosaic image to the \textsc{Gaia} catalog, and then registered the images in all of the other filters to the frame of the F444W image.
The mosaicked F444W image of the P330-E field is displayed in Figure~\ref{fig:01_map}.

Finally, we performed source extraction with the mosaicked F322W2 and F444W images using \textsc{SExtractor} v2.25.3 \citep{1996A&AS..117..393B} in the single-image detection mode down to $5\sigma$ detection levels.
The depths are $\sim$25.5 and 24.9\,AB mag in the F322W2 and F444W band, respectively, derived using the automatic Kron aperture \textred{and sky background measured from local annulus}.

\begin{figure*}
\centering
\includegraphics[width=0.9\linewidth]{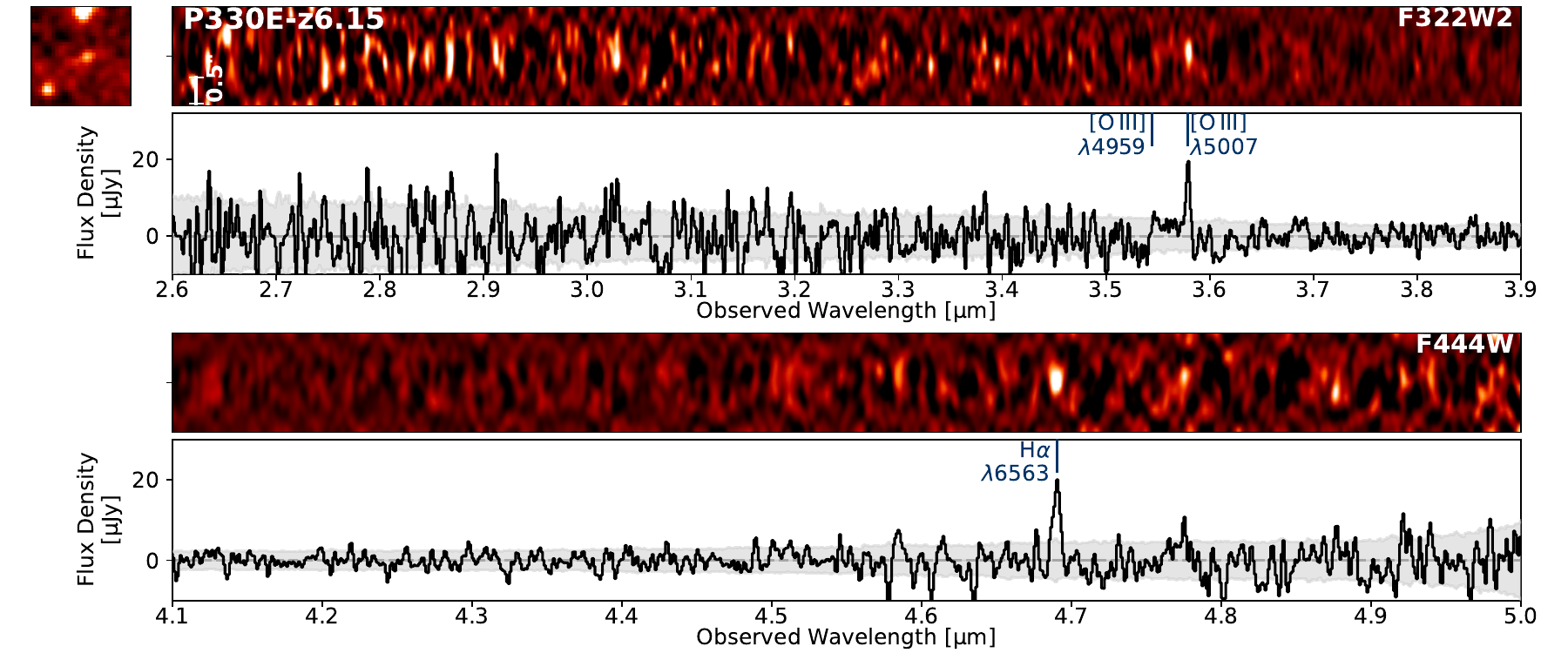}\\[1.5ex]
\includegraphics[width=0.9\linewidth]{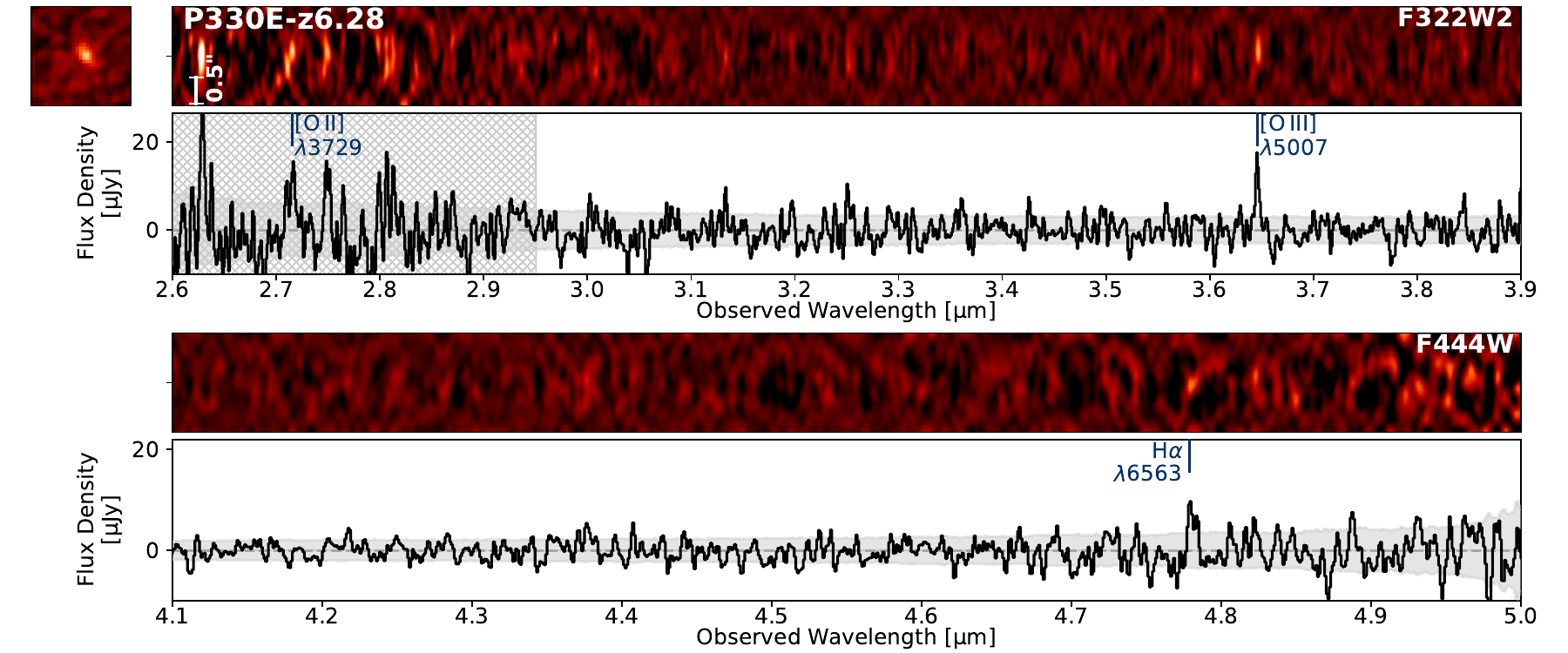}\\[1.5ex]
\includegraphics[width=0.9\linewidth]{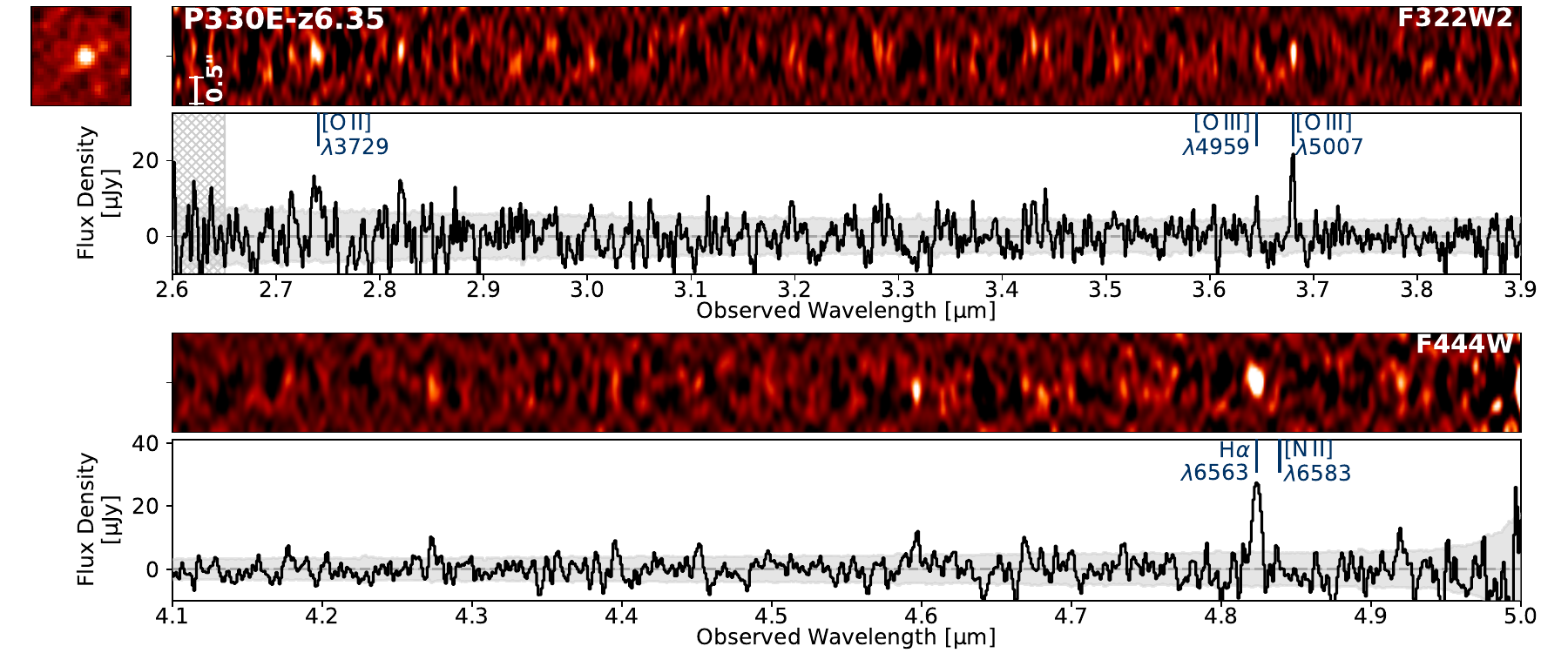}
\caption{Discovery image and 2D/1D spectra of three \oiii\ and \ha\ line emitters at $z=6.15$ (top), 6.28 (center) and 6.35 (bottom). 
In each plot, the NIRCam/F444W cutout image (size: 1\farcs9$\times$1\farcs9) is shown in the top-left panel.
Co-added, calibrated and background-subtracted 2D and 1D spectra in the F322W2 and F444W band are shown on the right.
Notable emission lines are labeled with dark blue lines.
1D error spectra are shown as the gray-shaded regions, and wavelength ranges with overlapping continuum contaminants are shown as the gray-hatched regions.
\textred{Note that the line widths of \oiii\ and \ha\ are consistent for each source, although they look different in the plot because of different intrinsic aspect ratios of the 2D spectra.}
}
\label{fig:02_spec}
\end{figure*}

\subsection{Grism Spectroscopy}
\label{ss:02b_grism}

As described in \citetalias{sun22}, the WFSS observations of the P330-E field were obtained with both the F322W2 (2.4--4.0\,\micron) and F444W (3.9--5.0\,\micron) filters on UT 2022 April 29.
The spectral resolution is $R\sim1600$ at around 4\,\micron, and the dispersion is $\sim 1$\,\si{nm.pixel^{-1}}.
In each band, the target was observed with four module (A/B)/grism (R/C; orthogonal) combinations (AR, AC, BR and BC) at four \textsc{intramodulex} primary dither positions, respectively.
With the five-group \textsc{bright1} readout pattern, the effective exposure time per integration was 96\,s.
Following \citetalias{sun22}, the last integration with the Module B Grism C in the F322W2 band (Observation \#108, Visit 001, Exposure 4) was not used in our analysis because of unstable guiding.
The maximum exposure time for a source is $\sim$25\,min in each band after combining integrations with all grisms. 
However, we also note that the actual effective exposure time could be shorter, depending on the wavelength of interest and the sky position of the source.
\textred{Therefore, the median exposure time in each band is $\sim$10\,min.}
We refer the reader to the JWST User Documentation\footnote{\href{https://jwst-docs.stsci.edu/jwst-near-infrared-camera/nircam-performance/nircam-wfss-field-of-view}{https://jwst-docs.stsci.edu/jwst-near-infrared-camera/nircam-performance/nircam-wfss-field-of-view}} \textred{and \citet{rieke23}} for further information.

The WFSS data were reduced to the level of Stage-1 (i.e., ``\textsc{\_rate}" files) with the same standard \textsc{jwst} calibration pipeline.
\textred{We applied a pixel-to-pixel flat-field correction using the imaging flat data obtained with the same filter and module. 
This is the same method as that we adopted for the F322W2/F444W grism flux calibration during the commissioning of the NIRCam/WFSS mode (also \citetalias{sun22}), because the large-scale grism flat-field calibration has not been complete at the time of writing. 
In the standard \textsc{jwst} pipeline, the step of grism background subtraction is performed by scaling the theoretical background to that observed in each individual integration. 
However, the accuracy of the current theoretical background is limited by the accuracy of the sky-background spectrum and the grism tracing/dispersion model (first and second order), neither of which have yet been fully characterized.
As in \citetalias{sun22}, we performed the 2D sky-background subtraction using the sigma-clipped median grism images. %, which were constructed from the obtained grism images.
}

Because the dithering of the telescope could introduce astrometric errors (i.e., the dithers might not exactly equal the commanded values), the WCS of each grism image was calibrated with the \textsc{Gaia} DR2 catalog by matching with the stars detected in the NIRCam SW images, which were taken simultaneously in the F212N band.
We note that such a registration relies on the internal alignment between NIRCam SW and LW instrument aperture and may introduce astrometric residuals. 
However, these residuals should be stable and have been included in the grism spectral tracing models.

The grism spectral tracing models, which give the relation between the spectral pixels $(x_s, y_s)$ and the direct-imaging position $(x_0, y_0)$, were constructed  using the spectral traces of point sources observed within the P330-E field.
The spectral tracing functions (e.g., $y_s(x_0, y_0, x_s)$ in Grism R) were constructed separately for the AR, AC, BR and BC module/grism combinations in the F322W2 and F444W filter. 
These functions include a third-degree polynomial of $x_0$, $y_0$ and a second-degree polynomial of $x_s$ to fit the strong curvature and field dependence of the spectral traces. 
With any given spectral pixel position in the dispersion direction (e.g., $x_{s}$ in Grism R), our spectral tracing model can predict the position along the perpendicular direction (e.g., $y_s$ in Grism R) with a root-mean-square (RMS) accuracy of 0.1--0.2\,pixel, i.e., 10--20\%\ of the RMS width of the point-source spectral trace.

The grism dispersion models, which give the relation between the spectral pixels along the dispersion direction (e.g., $x_{s}$ in Grism R) and $(x_0, y_0)$ and wavelength of interest ($\lambda$), were constructed using the emission-line spectra of IRAS\,05248--7007.
Up to eleven hydrogen recombination lines in Brackett ($n=4$), Pfund ($n=5$) and Humphreys ($n=6$) series were used for wavelength calibration in the F322W2 band, and up to eight lines were used in the F444W band, including hydrogen recombination lines and a \hei\ line at 4.296\,\micron\ (rest frame).
The dispersion functions (e.g., $x_s(x_0, y_0, \lambda)$ in Grism R) were constructed separately for the AR, AC, BR and BC module/grism combinations using the F322W2 and F444W data simultaneously.
These functions include a second-degree polynomial of $x_0$, $y_0$ and a third-degree polynomial of $\lambda$ to perform wavelength calibration with field dependence.
Our dispersion model can predict the position of a spectral feature at a wavelength of $\lambda_{s}$ along the dispersion direction (e.g., $x_{s}$ in Grism R) with an RMS accuracy of 0.2\,pixel, i.e., $\sim$\,10\%\ of the two-pixel resolution element.
 
We also used the spectra of P330-E to construct the flux calibration functions in the F322W2 and F444W filters. 
Using the grism tracing and dispersion models described above, we extracted the spectra of the standard star using box apertures with a height of $D = 20$\,pixels (1\farcs26).  The corresponding aperture loss was found to be small (2--3\%) when compared with the extracted spectra with a larger aperture ($D=100$\,pixels).
This loss was corrected in the analyses below.
The flux calibration functions, i.e., conversion factors from count rate (unit: DN/s) to flux density (unit: mJy) as functions of wavelength, were constructed using all of the available integrations for four grisms taken in the F322W2 and F444W filters, respectively, with a wavelength step of 0.005\,\micron\ ($\sim5$\,pixels).
Through the comparisons of extracted P330-E spectra in different exposures and different locations on the detector, we find that the variations of count rates are only 1--2\%\ when the spectra are binned to 0.005\,\micron.
We therefore conclude that the accuracy of flux calibration is dominated by the systematic uncertainty ($\sim$2\% from standard star; \citealt{gordon22}) and flat-field error ($\sim3$\% for pixel variation).

\textred{The grism spectral tracing, dispersion and flux calibration models derived from the commissioning data have been made available online\footnote{\href{https://github.com/npirzkal/GRISM_NIRCAM/}{https://github.com/npirzkal/GRISM\_NIRCAM/}}.}
With the derived source catalog in Section~\ref{ss:02a_img} and models described above, we conducted 2D spectral extraction, wavelength and flux calibration on the flat-fielded WFSS data, and combined the extracted 2D spectra taken with Grism R, C and both.
These were performed for $\sim$3000 sources detected in the F322W2 and F444W image as described in \citetalias{sun22}.

\section{Results} \label{sec:03_res}

\begin{figure*}[!t]
% \centering
\begin{minipage}[l]{0.49\textwidth}
\includegraphics[width=\linewidth]{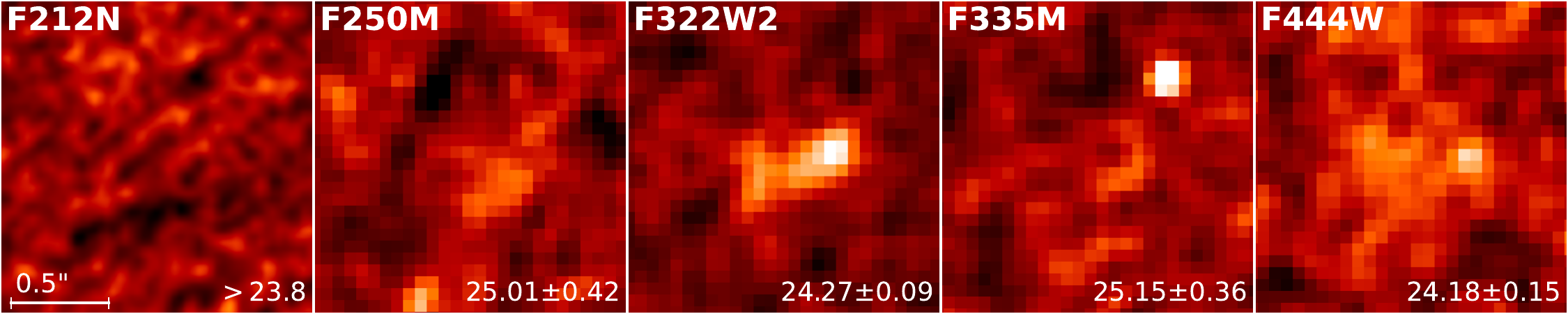}
\includegraphics[width=\linewidth]{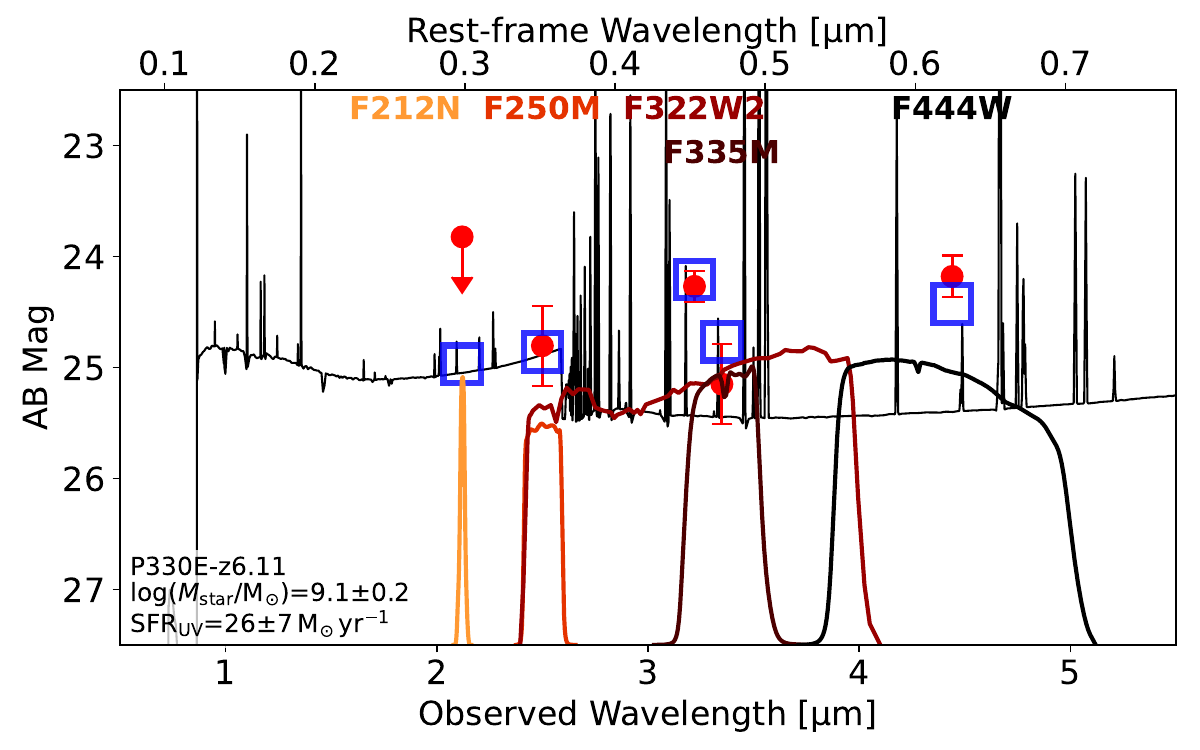}
\end{minipage}\hspace{6pt}
\begin{minipage}[l]{0.49\textwidth}
\includegraphics[width=\linewidth]{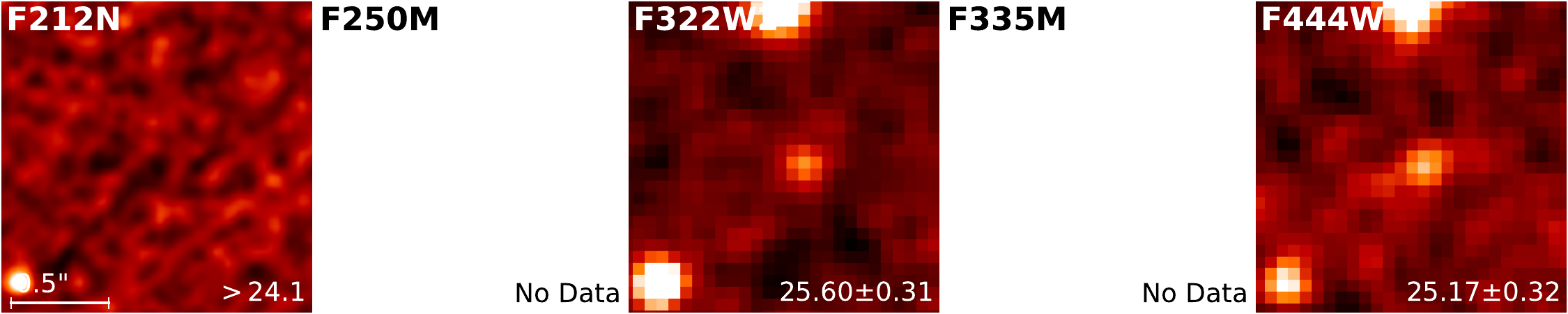}
\includegraphics[width=\linewidth]{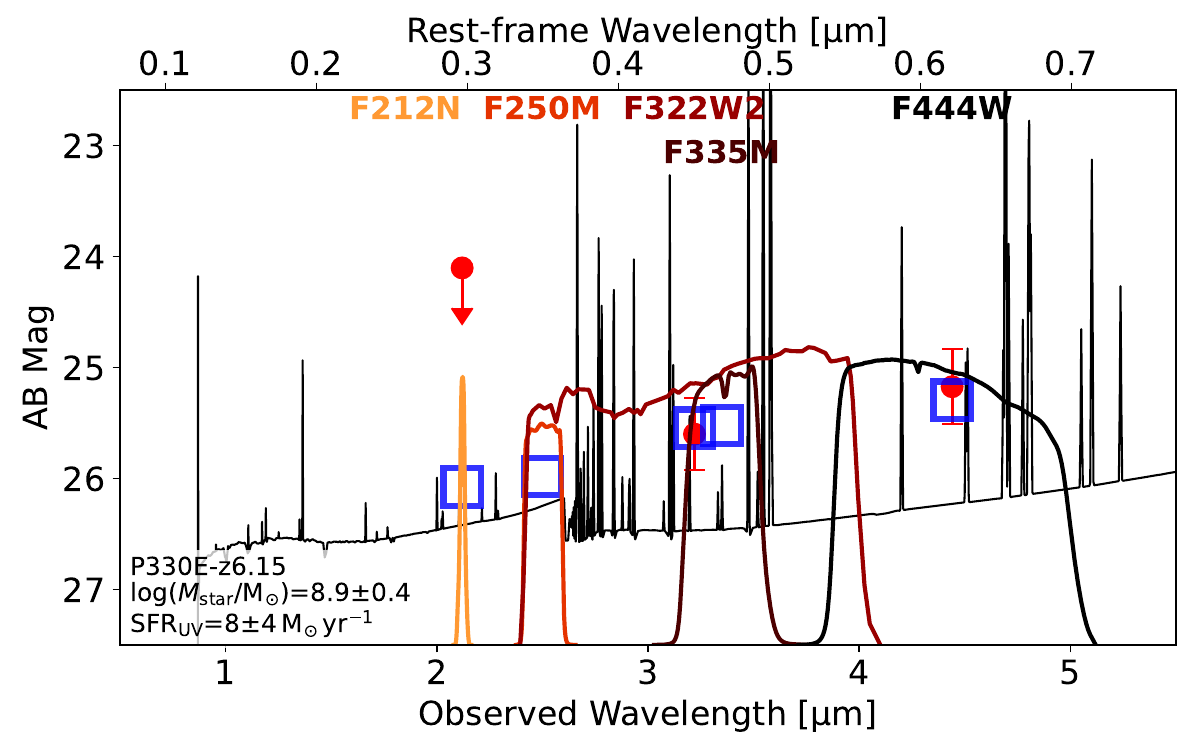}
\end{minipage}\\[1.5ex]
\begin{minipage}[r]{0.49\textwidth}
\includegraphics[width=\linewidth]{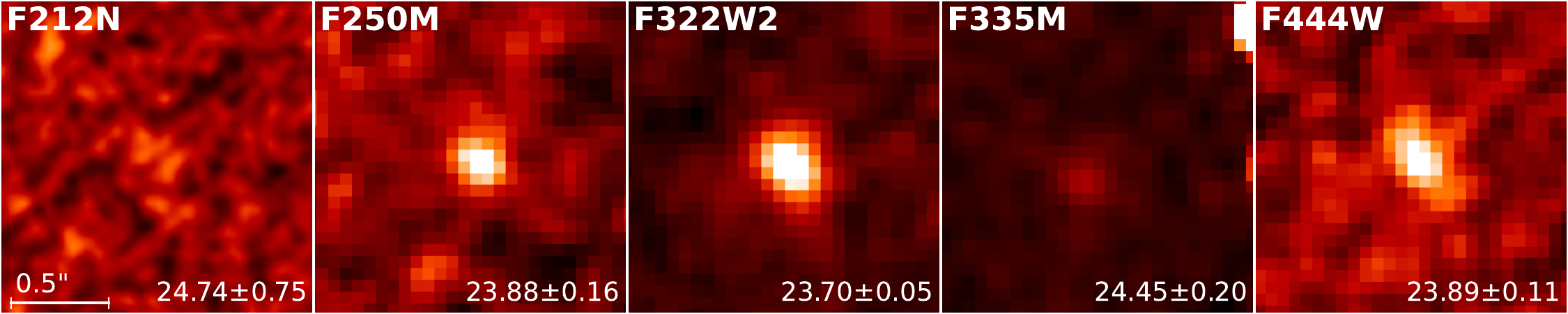}
\includegraphics[width=\linewidth]{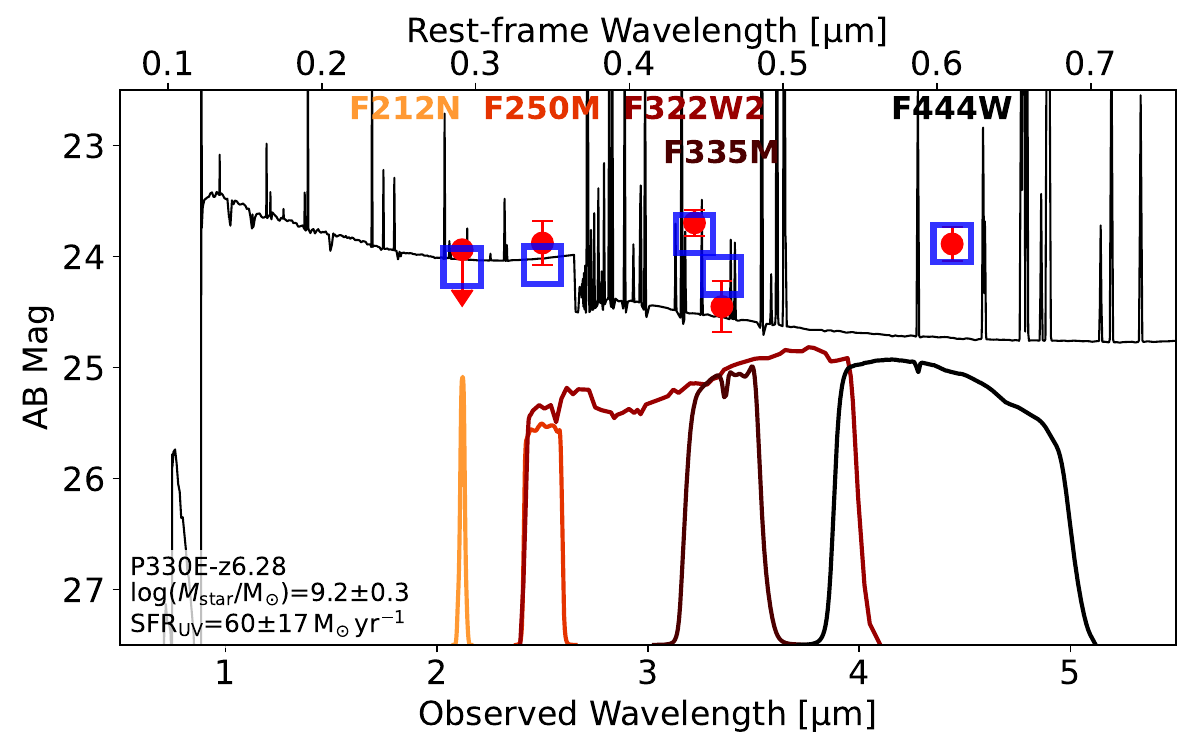}
\end{minipage}
\begin{minipage}[l]{0.49\textwidth}
\includegraphics[width=\linewidth]{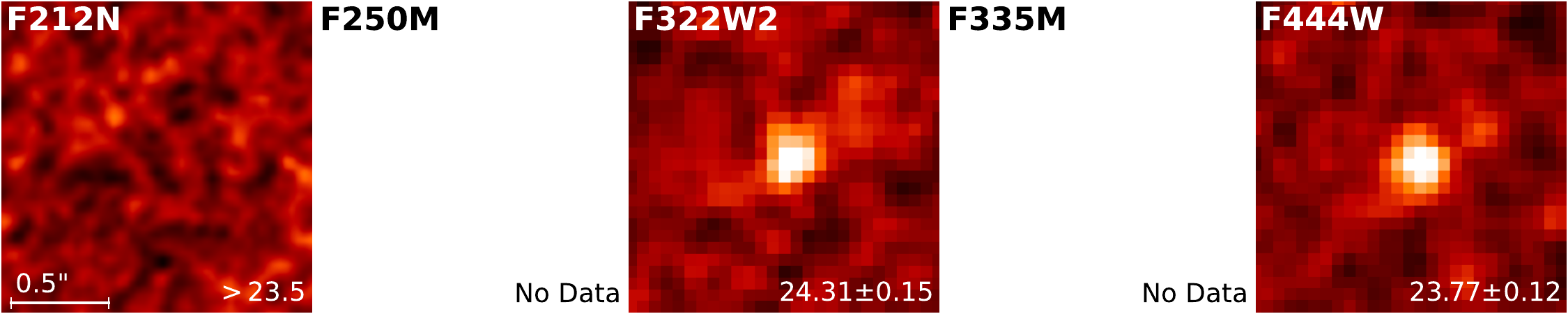}
\includegraphics[width=\linewidth]{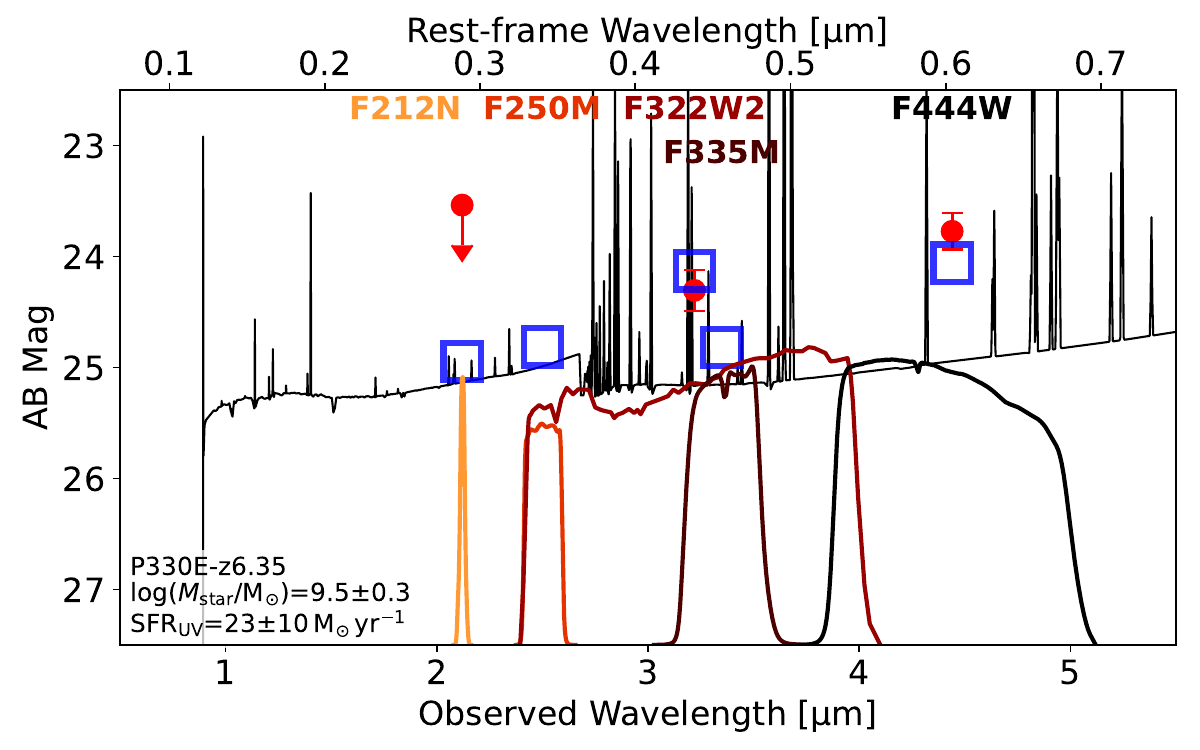}
\end{minipage}
% \begin{minipage}[r]{0.49\textwidth}
% \includegraphics[width=\linewidth]{figure/cutout_placeholder.pdf}
% \end{minipage}
\caption{JWST/NIRCam multiple-wavelength cutout images and SED models of the four sources presented in this work. 
% The images and SED model of the $z=6.11$ source have been presented in \citetalias{sun22} \textred{are updated}.
In each panel we show F212N, F250M, F322W2, F335M and F444W cutout images on the top. Measured source brightnesses are shown in the lower-right corner of each cutout image (unit: AB\,mag; $3\sigma$ upper limit for non-detection).
These measurements are shown as \textred{filled} red circles in the SED plot.
Best-fit SED models obtained with \textsc{cigale} are shown as black curve, and best-fit source brightnesses in all filters are shown as open blue squares.
The transmission curves of all used filters are also shown for comparison.
Derived stellar mass and UV-based SFR are shown in the lower-left corner of each SED plot.
}
\label{fig:03_sed}
\end{figure*}

\subsection{Discoveries of \ha$+$\oiii\ Emitters at $z>6$}
\label{ss:03a_spec}

We limited the search for line emitters at $z>6$ to a survey area that could yield detections of both \oiii\,$\lambda$5007 (in the F322W2 band, observed wavelength $\lambda_\mathrm{obs} > 3.5$\,\micron) and \ha\ lines (in the F444W band; $\lambda_\mathrm{obs} > 4.6$\,\micron).
These two emission lines are expected to be the most luminous lines of star-forming galaxies in the rest-frame optical, and the detection of these two lines can securely determine the redshift.
The effective survey area, up to 14.4\,\si{arcmin^2}, depends on the redshift and expected line luminosities, which will be further discussed in Appendix~\ref{apd:01}.

% {\color{red} (Much of the text in this paragraph is tecnical in nature, e.g., extraction of 1D spectra, and therefore should be presented in Sec 2.)}
Line emitters were identified through visual inspection of 2D grism spectral images, 1D spectra and 2D direct images taken in both the F322W2 and F444W bands.
Because the underlying stellar continuum emission of EoR galaxies is not expected to be detectable with the shallow grism data in the combined 2D spectral images, we subtracted the \textred{median-filtered 2D background.
The 1D spectra were then optimally extracted \citep{horne86} based on the Gaussian models of emission-line profiles in the spatial direction.
The 1D uncertainty spectra were generated using the error extension of the 2D spectra, and we confirm that they are consistent with the RMS of 1D scientific spectra.
}
% 1D spectra were extracted uniformly using box apertures with a height of $D = 7$\,pixel (0\farcs44), which typically maximized the S/N of lines with a FWHM of $\sim$\,5\,pixel (0\farcs3) in this study.
% If a line was detected, an aperture correction factor would be derived through the modeling of line profiles in the vertical direction.
All of the line profiles in collapsed 1D spectra were modeled with Gaussian profiles, and potential \oiii\ and \ha\ emission lines with a signal-to-noise ratio (S/N) that was less than 3 were discarded from further analysis.

Among $\sim$3000 direct-imaging sources detected in either the F322W2 or F444W band, $\sim$10\%\ of them yielded potential emission-line detections (see the $z=4.4$ \oiii\ and \ha\ line emitter in the JWST commissioning report; \citealt{rigby22}). 
Four sources were confirmed at $z>6$ after careful inspection, including the $z=6.112$ source reported in \citetalias{sun22} and three new sources at $z=6.146$, 6.279 and 6.348.
All of these sources yielded secure \oiii\,$\lambda$5007 ($>5\sigma$) and \ha\ ($>3\sigma$) line detections in the coadded spectra, which could be detected in both the R and C grism data separately as long as the lines are within the wavelength coverage.
These galaxies are referred to as P330E-z6.11, P330E-z6.15, P330E-z6.28 and P330E-z6.35 hereafter based on the ID of the primary target in the field and their redshifts.

Figure~\ref{fig:01_map} shows the locations of these $z>6$ emission-line galaxies in the mosaicked NIRCam F444W map.
The cutout direct images, 2D spectra and 1D spectra are shown in Figure~\ref{fig:02_spec} with notable emission lines labeled.
Among them, P330E-z6.11 and z6.28 were always observed with the flux calibrator in the same module of NIRCam, and therefore their effective integration time is $\sim$20\,min with each filter, which is roughly twice that of the other two sources.
We also note that a few other suspicious lines can be identified in the 2D spectra, but most of them were only detected in the spectra produced by either the R or C grism, and therefore their associations with the sources are doubtful.

We modeled the properties of detectable emission lines (in increasing order of wavelength: \oii\,$\lambda$3727, \hb, \oiii\,$\lambda\lambda$4959, 5007, \ha\ and \nii\,$\lambda$6583) using Gaussian profiles.
Among them, \hb+\oiii\ line profiles were fitted simultaneously, and the line centers were controlled by the redshift parameter. 
Given the low significance of the detection, the \hb\ and \oiii\,$\lambda$4959 line FWHMs were fixed to be the same as those of the \oiii\,$\lambda$5007 lines.
\ha$+$\nii\,$\lambda$6583 lines were modeled in the same way.  
The best-fit redshift parameters are consistent with those derived from \hb+\oiii\ fitting within $\Delta z \sim 0.001$, demonstrating the accuracy of the wavelength calibration.
We fitted the flux of the \oii\,$\lambda\lambda$3726, 3729 lines using a single Gaussian profile at $\lambda$3727 because the doublet cannot be resolved. 
In this fit, the redshift parameter was fixed and the line FWHM was set to be identical to \textred{the average FWHM of the \ha\ and \oiii\ lines} to mitigate the artificial broadening and flux-boosting effect because of low significance of the line detections (see discussion in Appendix~\ref{apd:01}).

All of the line flux measurements are presented in Table~\ref{tab:01_prop}.
In addition to the firm detections of \oiii\,$\lambda$5007 and \ha\ lines, \oii\,$\lambda$3727 lines were detected in P330E-z6.28/z6.35 at $\sim 2.6\sigma$, which appear to be more luminous than their \oiii\,$\lambda$5007 lines.
% {\color{red} (Remove the following line because it is an interpretation.)}
% \st{This may indicate a relatively low ionizing and/or metal-enriched environment within the interstellar medium (ISM) of these two galaxies (further discussed in Section~}\ref{ss:04c_metal}).
\hb\ was not detected for the three new sources presented here, while the $3\sigma$ upper limits of line fluxes are consistent with or higher than those estimated from \ha\ line fluxes assuming Case B recombination with a typical electron temperature of $T_\mathrm{e}=10^4$\,K \citep[\ha/\hb=2.86;][]{osterbrock06}.
The \oiii\,$\lambda$4959 lines were also tentatively detected for all sources (\textred{up to $3.7\sigma$}), and the line ratios to \oiii\,$\lambda$5007 line are consistent with the theoretical ratio of 1/3.
\textred{Finally, the significance of \nii\,$\lambda$6583 line is $<2\sigma$ for all sources in our sample.}
% Finally, \nii\,$\lambda$6583 line was tentatively detected for P330E-z6.35 ($1.5\sigma$), which is also seen for P330E-z6.11 as reported in \citetalias{sun22}.

\begin{table*}[!t]
\caption{Summary of the properties of $z>6$ \ha\ and \oiii\ emitters in this work.}
\centering
\begin{tabular}{lrrrr}
\hline\hline & P330E-z6.11  & P330E-z6.15  & P330E-z6.28  & P330E-z6.35                \\\hline
R.A.         & 16:31:34.46  & 16:31:21.79  & 16:31:37.02  & 16:31:47.34                \\
Decl.        & +30:08:10.5  & +30:10:18.4  & +30:08:28.1  & +30:06:34.2                \\
Redshift     & 6.112$\pm$0.001 & 6.145$\pm$0.001 & 6.280$\pm$0.001 & 6.348$\pm$0.001    \\
\hline\hline Photometric Properties                        \\\hline
 F212N [AB mag]  &        $>$23.8 &        $>$24.1 & 24.74$\pm$0.75 &        $>$23.5 \\
 F250M [AB mag]  & 25.01$\pm$0.42 &       \nodata  & 23.88$\pm$0.16 &       \nodata  \\
F322W2 [AB mag]  & 24.27$\pm$0.09 & 25.60$\pm$0.31 & 23.70$\pm$0.05 & 24.31$\pm$0.15 \\
 F335M [AB mag]  & 25.15$\pm$0.36 &       \nodata  & 24.45$\pm$0.20 &       \nodata  \\
 F444W [AB mag]  & 24.18$\pm$0.15 & 25.17$\pm$0.32 & 23.89$\pm$0.11 & 23.77$\pm$0.12 \\
\hline\hline Spectroscopic Properties                      \\\hline
$f$(\oii\,$\lambda$3727) [\si{10^{-18}\,erg.s^{-1}.cm^{-2}}]  &      $<$30.9 &      $<$52.3 & 22.2$\pm$9.5 & 30.7$\pm$10.9 \\
$f$(\hb) [\si{10^{-18}\,erg.s^{-1}.cm^{-2}}]                       &  7.1$\pm$2.8 &      $<$15.6 &       $<$5.2 &       $<$9.6 \\
$f$(\oiii\,$\lambda$4959) [\si{10^{-18}\,erg.s^{-1}.cm^{-2}}]      &  9.9$\pm$2.7 &  7.3$\pm$3.6 &  2.1$\pm$1.7 &  8.7$\pm$3.1 \\
$f$(\oiii\,$\lambda$5007) [\si{10^{-18}\,erg.s^{-1}.cm^{-2}}]      & 34.8$\pm$3.2 & 24.8$\pm$4.1 & 13.8$\pm$2.1 & 20.9$\pm$3.6 \\
$f$(\ha) [\si{10^{-18}\,erg.s^{-1}.cm^{-2}}]                       & 14.7$\pm$2.6 & 16.9$\pm$3.2 &  7.6$\pm$2.3 & 26.5$\pm$3.8 \\
$f$(\nii\,$\lambda$6583) [\si{10^{-18}\,erg.s^{-1}.cm^{-2}}]       &       $<$6.3 &       $<$7.0 &       $<$5.8 &  3.6$\pm$2.9 \\
EW(\oii\,$\lambda$3727) [\AA]   &       $<$213 &     \nodata  &    97$\pm$45 &  291$\pm$114 \\
EW(\hb) [\AA]                        &    70$\pm$29 &       $<$729 &        $<$35 &        $<$94 \\
EW(\oiii\,$\lambda$4959) [\AA]       &   100$\pm$30 &  344$\pm$254 &    14$\pm$12 &    85$\pm$33 \\
EW(\oiii\,$\lambda$5007) [\AA]       &   359$\pm$60 & 1165$\pm$670 &    97$\pm$22 &   205$\pm$49 \\
EW(\ha) [\AA]                        &   221$\pm$50 &  841$\pm$489 &    84$\pm$29 &   267$\pm$59 \\
EW(\nii\,$\lambda$6583) [\AA]        &        $<$96 &       $<$353 &        $<$63 &    36$\pm$30 \\
\hline\hline Physical Properties                      \\\hline
$\log[M_\mathrm{star}/\msun]$     & 9.1$\pm$0.2 & 8.9$\pm$0.4 & 9.2$\pm$0.3 & 9.5$\pm$0.3   \\
SFR(\ha) [\smpy]                  & 32$\pm$5 & 38$\pm$7 & 18$\pm$5 & 64$\pm$9    \\
SFR(SED,UV) [\smpy]               & 25$\pm$7 & 7$\pm$4 & 60$\pm$17 & 22$\pm$10    \\
$\log[\xi_\mathrm{ion}/(\mathrm{erg}^{-1}\mathrm{Hz})]$ & 25.2$\pm$0.1 & 25.8$\pm$0.3 & 24.6$\pm$0.2 & 25.5$\pm$0.2    \\
12 + $\log$(O/H)   & 8.2$\pm$0.2 & 8.3$\pm$0.2 & 8.2$\pm$0.3 & 8.4$\pm$0.2    \\
\hline
\end{tabular}
\tablecomments{Properties of P330E-z6.11 were reported in \citet{sun22}, \textred{but here we update all measurements with the latest flux calibration.}
}
\label{tab:01_prop}
\end{table*}

\subsection{Photometry and Line Equivalent Widths}
\label{ss:03b_phot}

% {\color{red} (Again, much of the text in this paragraph is tecnical in nature, e.g., photometry procedure, and therefore should be presented in Sec 2.)}
Similar to \citetalias{sun22}, we performed aperture photometry of all $z>6$ line-emitting galaxies in F212N, F250M, F322W2, F335M and F444W band. 
Unlike P330E-z6.11 that has a clear two-component structure, the three new sources found in this work were detected as single-component systems and their angular sizes are generally compact (FWHM\,$\lesssim$\,0\farcs25).
We adopted a conservative circular aperture of $r=0\farcs45$ using \textsc{photutils} \citep{photutils} to avoid missing extended components and minimize the aperture loss, similar to that used in \citetalias{sun22}.
\textred{With such an aperture radius, the variation of the encircled light fractions of point spread functions at 2--5\,\micron\ is $\sim$3\%, much smaller than the photometric uncertainty.}
\textred{We also modeled the F444W surface brightness profiles of sources in our sample with the 2D Gaussian model, and we confirmed that the aperture loss is negligible.}
% Through the modeling of 4.4-\micron\ circularized surface brightness profiles of all sources in our sample, we confirmed that the aperture correction is negligible.
We subtracted the sky background using the median of sigma-clipped local annulus, and computed the photometric uncertainty using the RMS of that.
Broad-band photometry is summarized in Table~\ref{tab:01_prop}.

All sources were detected in the broad F322W2 and F444W bands, but remained undetected within the narrow F212N filter.
For P330E-z6.28 with the F250M and F335M coverage, we were able to detect its rest-frame UV continuum (\textred{23.88$\pm$0.16}\,AB mag in the F250M band) and optical stellar continuum (\textred{24.45$\pm$0.20}\,AB mag in the F335M band).
At this redshift, no strong emission line is expected in the wavelength range of the F250M and F335M filters, but the flux densities measured with the broad F322W2 and F444W filters could be boosted by strong \oiii\ and \ha\ lines, respectively.
Indeed, we find a blue F250M--F335M color of \textred{$-0.57\pm0.26$} and an F322W2--F335M excess of \textred{$0.75\pm0.21$}\,mag, both suggesting the presence of strong nebular emission lines (e.g., \oiii) and \textred{potentially} nebular continuum blueward of the Balmer break ($\lambda_\mathrm{rest} < 3640$\,\AA).

With the continuum photometry, we computed the line equivalent widths (EWs).
To estimate the underlying continuum flux density, we subtracted all measured line fluxes within the passband for all broad-band photometric measurements redward of the Balmer break (F322W2, F444W and F335M if available).
The continuum flux densities were then modeled with a power-law function $f_\mathrm{\nu} \propto \lambda ^ \alpha$.
We then computed the line EW using the underlying continuum strength estimated at the line wavelength.
All of the line EWs are reported in Table~\ref{tab:01_prop}.
However, we also note that the broad-band flux density could be slightly overestimated because of the contribution from undetected faint emission lines that may not have been properly subtracted, which could potentially lead to underestimates in line EWs. 
The Balmer discontinuity can also introduce errors into the continuum flux density determination in the F322W2 band. 
%, as is seen for P330E-z6.28.

\section{Discussion \RomanNumeralCaps{1}: Physical Properties} \label{sec:04_disc}

\subsection{SED modeling}
\label{ss:04a_sed}

% {\color{red} (Modeling is interpretation by definition, and therefore this 3.3 doesn't belong to the Results section.  Move this to the Discussion section instead.)}

\textred{To derive the physical properties of galaxies in our sample}, we perform SED modeling with \textsc{cigale} \citep{boquien19}.
Broad-band photometry and the EWs of \oiii\ and \ha\ lines are included as constraints. 
\oii, \hb\ and \nii\ line EWs are not included given the general low significance of detections.
Similar to that in \citetalias{sun22}, we assume a commonly used delayed star-formation history (SFH; \texttt{sfhdelayed}), in which $\mathrm{SFR}(t) \propto t \exp(-t/\tau)$ and $\tau$ is the peak time of SFH. 
An optional late starburst is allowed in the last 1--5\,Myr, which can contribute to 0--80\%\ of the total stellar mass.
We use \citet{bc03} stellar population synthesis models.
We also allow a metallicity range of 0.2\,\si{Z_\odot} to \si{Z_\odot}, a broad ionization parameter ($\log U$) range of $-1.0$ to $-3.5$.
We adopt the \citet{calzetti00} attenuation curve and the color excess of the nebular lines is allowed between $E(B-V)_\mathrm{line} = 0 - 1$.

The best-fit SED models are shown in Figure~\ref{fig:03_sed}.
\textred{The best-fit SEDs are rich in emission lines, indicating that the galaxies are young and star-forming. However, most of the lines are too faint to be detected in the grism spectra.}.
Similar to the results in \citetalias{sun22}, the best-fit SFH models typically invoke both a young (1--2\,Myr) and old ($\sim300$\,Myr) stellar populations.
Roughly half of the stellar masses in all galaxies are produced by the most recent starburst, which is consistent with the presence of strong nebular emission lines. 
\textred{We note that this result is driven by the large emission-line EWs. 
If the fitting is performed without line EWs, then the amount of recent star formation will decrease sharply.}

The median mass-weighted stellar age of sources in our sample is \textred{$74\pm50$}\,Myr, and the median stellar mass ($M_\mathrm{star}$) is \textred{$1.4_{-0.5}^{+0.7}\times 10^9$}\,\msun.
Dust attenuation is negligible for P330E-z6.11/6.28, which is likely not the case for P330E-z6.15 (\textred{$A_V=1.1\pm0.7$} for stellar continuum) and P330E-z6.35 (\textred{$A_V=1.4\pm0.7$}) as indicated by their red F322W2--F444W colors even after the subtraction of \oiii\ and \ha\ lines.
However, we also note that the constraints on the dust attenuation and stellar masses are not tight, given that three of the sources were only detected in the two broad LW bands.

The derived physical properties of galaxies in our sample are also presented in Table~\ref{tab:01_prop}.
The physical properties of P330E-z6.11 have been reported in \citetalias{sun22}, \textred{and we have updated our results with the latest flux calibration.
The differences from those in \citetalias{sun22} are small,}
% and we confirm that those values remain valid with the new uniform data reduction, photometric/spectroscopic measurements and SED modeling, as the differences are 
within $1\sigma$ significance level.
Following \citetalias{sun22}, we also infer UV SFRs for galaxies in our sample from the best-fit SED, and the median is \textred{$24\pm11$}\,\smpy.
This is smaller than the median of \ha-based SFR (\textred{$35\pm9$}\,\smpy) assuming the conversion in \citet[uncorrected for dust attenuation]{ke12}.
This comparison likely indicates (\romannumeral1) a bursting nature because \ha\ \textred{is more sensitive to the most} recent star formation than UV continuum (see \citealt{ke12}),  (\romannumeral2) a higher ionizing photon production efficiency ($\xi_\mathrm{ion}$), \textred{lower metallicity and thus smaller \ha/SFR conversion factor \citep[e.g.,][]{charlot02,brinchmann04}} at high redshift when compared with the local Universe, and  (\romannumeral3) a higher dust extinction in the rest-frame UV than with the \ha\ line (see further discussions in Section~\ref{ss:04e_ha}).

% \textbf{(move below to discussion?)}

\begin{figure*}[!t]
\centering
\includegraphics[width=0.49\linewidth]{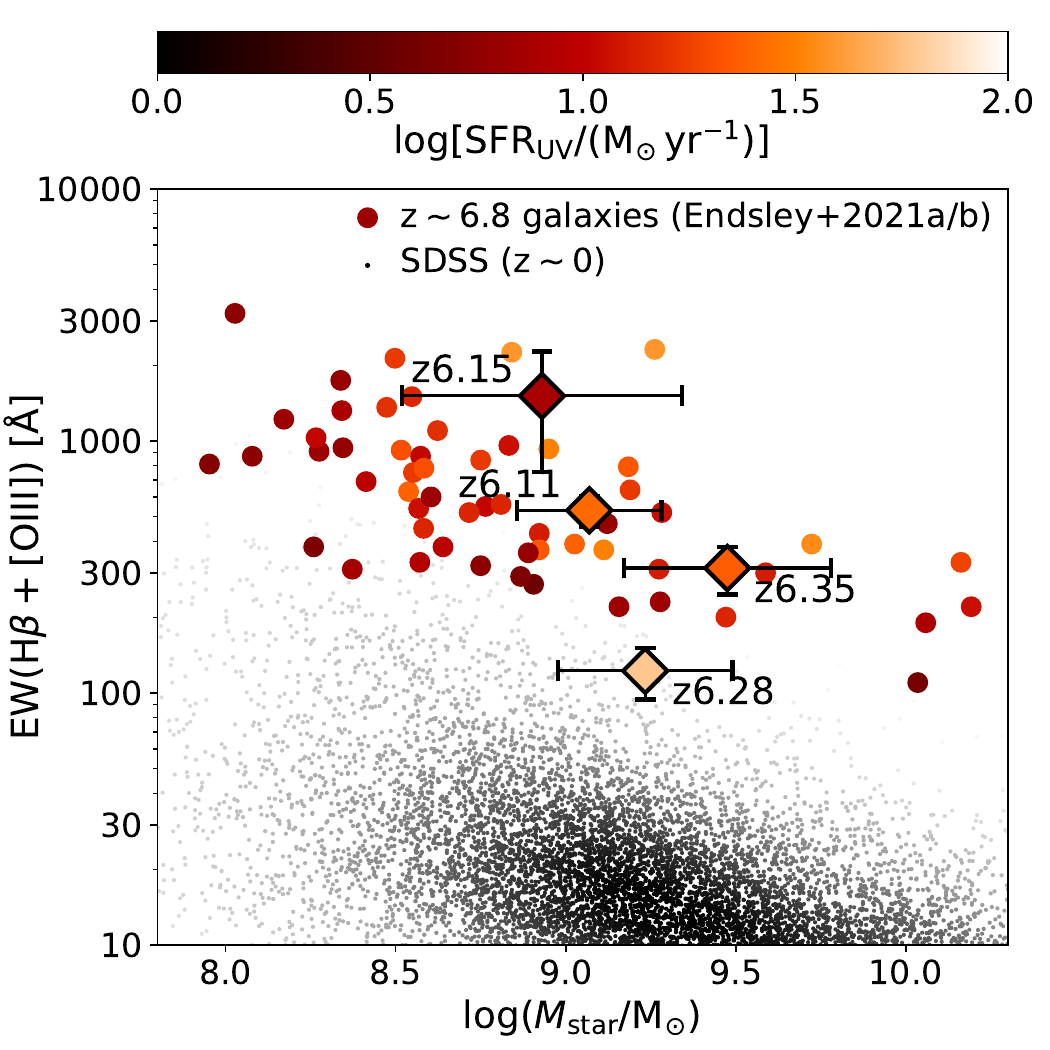}
\includegraphics[width=0.49\linewidth]{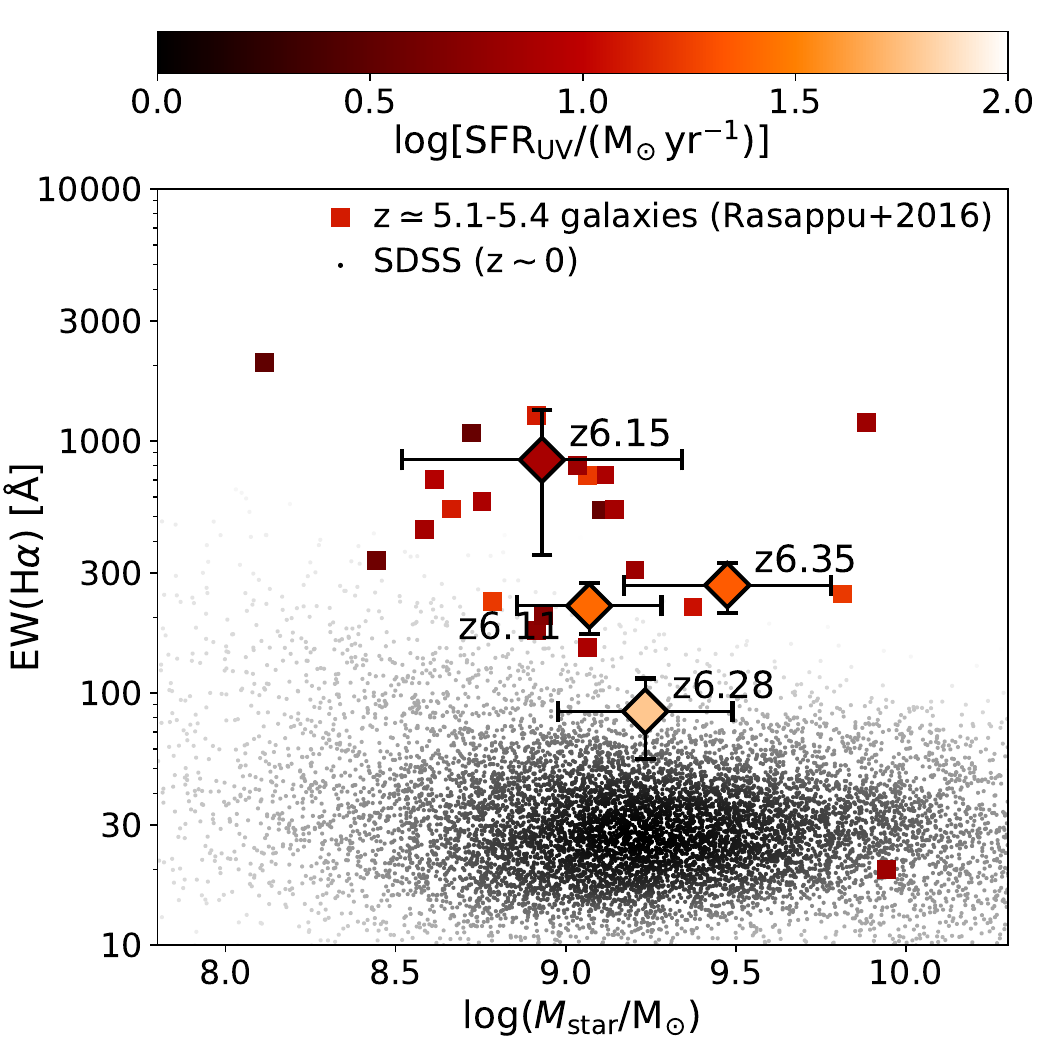}
\caption{Left: \hb$+$\oiii\ EWs versus stellar masses of sources in our sample (diamonds), compared with galaxies at $z\sim6.8$ whose line EWs are inferred from Spitzer/IRAC [3.6]--[4.5]\,\micron\ colors (circles; \citealt{endsley21b,endsley21a}). 
Right: \ha\ EWs versus stellar masses of sources in our sample (diamonds), compared with galaxies at $z\simeq5.1-5.4$ whose line EWs are also inferred from IRAC colors (squares; \citealt{rasappu16}). 
\textred{In both panels,  all high-redshift sources are color-coded by their UV-based SFRs.
SDSS-selected galaxies in the local Universe are also shown as dots for comparison.}
}
\label{fig:04_ew_ms}
\end{figure*}

\begin{figure}[!t]
\centering
\includegraphics[width=\linewidth]{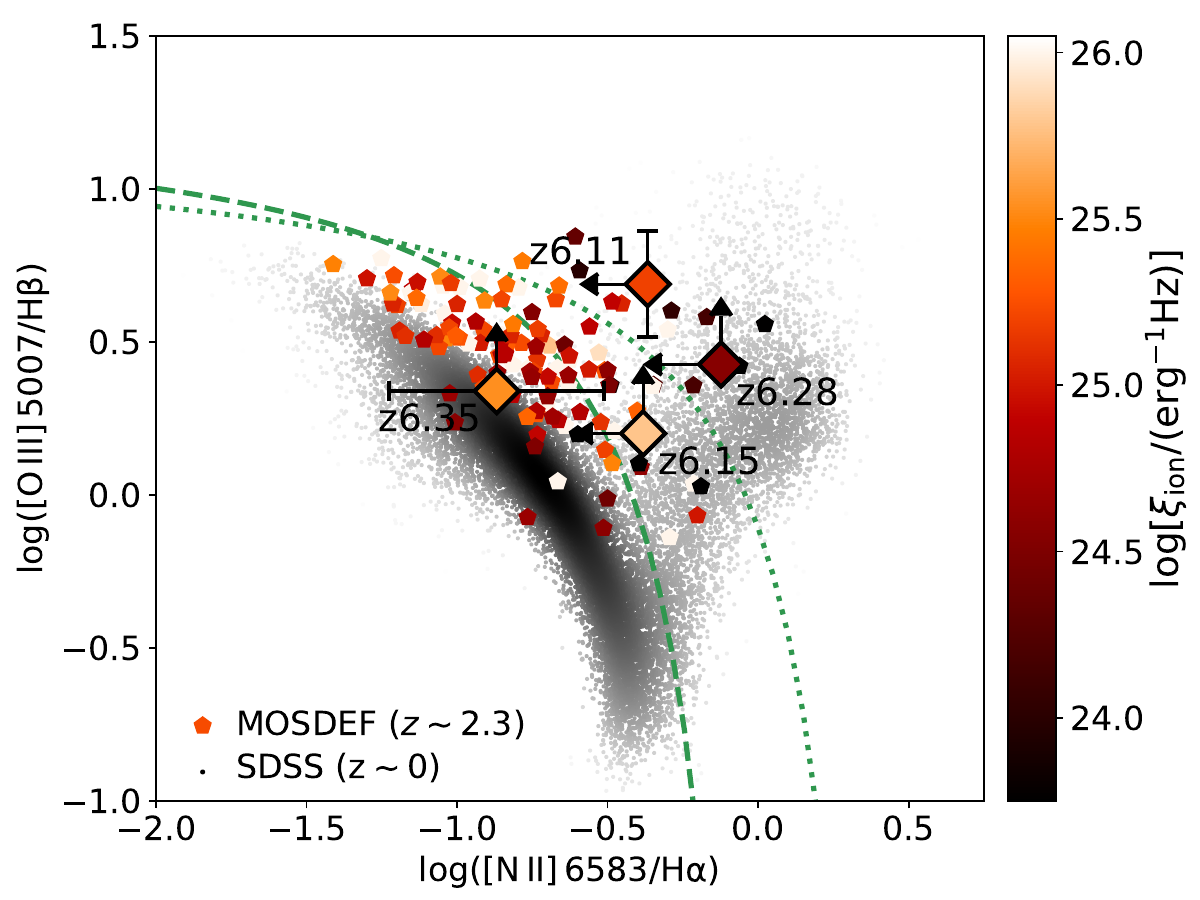}
\caption{\oiii/\hb--\nii/\ha\ BPT diagram of $z>6$ galaxies in our sample (diamonds), compared with those of $z\sim2.3$ emission-line galaxies in MOSDEF sample \citep[pentagons;][]{kriek15,reddy15,shapley15}.
Galaxies in both samples are color-coded by their ionizing photon production efficiencies, and likely occupy the same parameter space.
SDSS-selected galaxies \textred{and  AGN} in the local Universe are shown as dots in background, color-coded by their number densities in the parameter space.
The dotted green line is the so-called ``maximum-starburst'' line in \citet{kewley01}, and the dashed green line is the canonical  AGN/star-forming galaxy boundary in \citet{kauffmann03}.
}
\label{fig:05_bpt}
\end{figure}

\begin{figure}[!t]
\centering
\includegraphics[width=\linewidth]{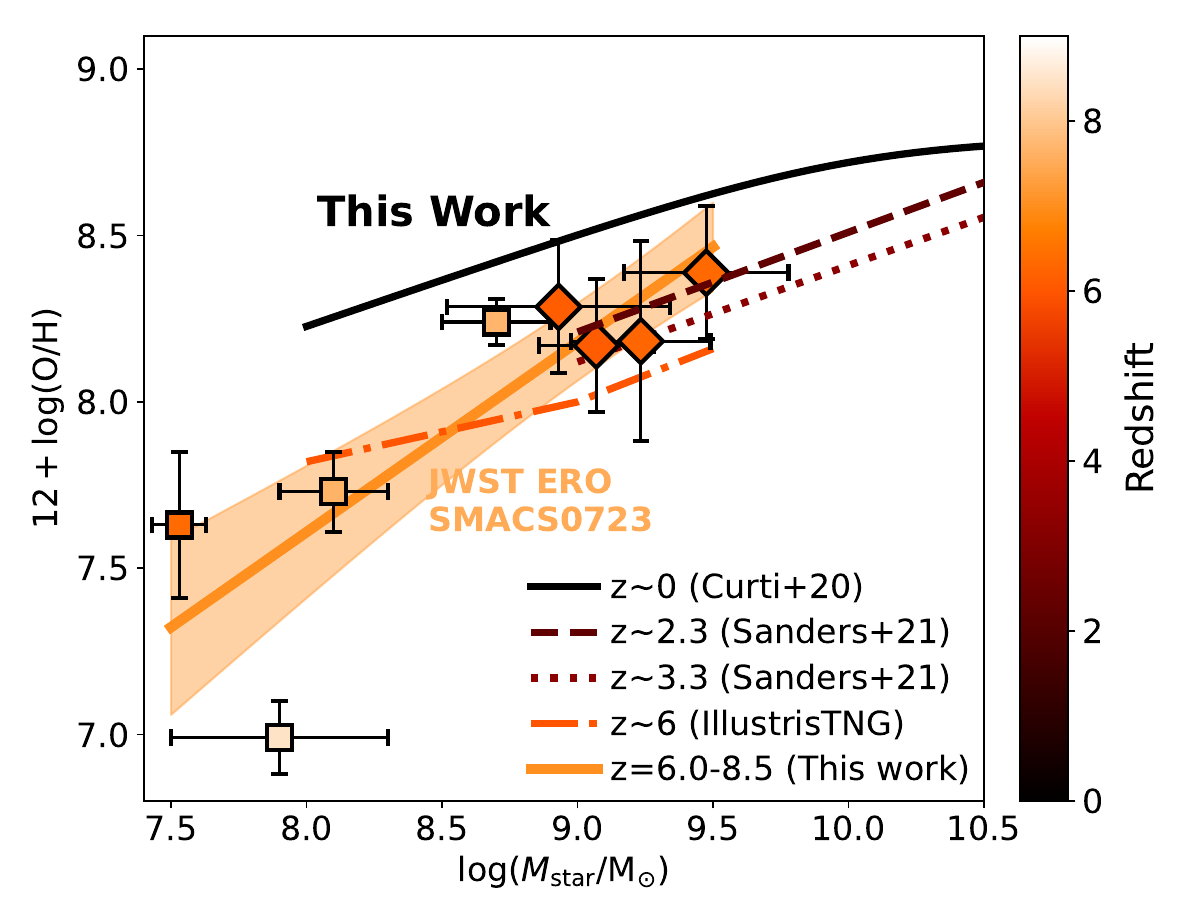}
\caption{Gas-phase metallicity versus stellar mass of $z>6$ galaxies in our sample (diamonds; measured using various strong line ratios calibrated by \citealt{bian18}), compared with those obtained with JWST/NIRSpec early-release observations in SMACSJ0723 field (squares; \citealt{curti22}, \citealt{taylor22}) measured using the direct $T_\mathrm{e}$ method.
The best-fit mass-metallicity relation of $z>6$ galaxies is shown as solid orange line, with $1\sigma$ uncertainty shown as shallower filled region.
We also compare our measurements with those of SDSS galaxies in the local Universe \citep{curti20}, and $z\sim2.3$ and 3.3 galaxies in the MOSDEF sample (\citealt{sanders21}; also based on \citealt{bian18} calibration).
Mass-metallicity relation at $z\sim6$ in the IllustrisTNG simulation \citep{torrey19} is also shown for comparisons (dash-dotted orange-red line).
All samples and relations shown in this plot are color-coded by the redshifts.
}
\label{fig:06_mzr}
\end{figure}

\subsection{Comparison of Line EWs with $z\simeq 5 - 7$ galaxies}
\label{ss:04b_ew}

\textred{Before the launch of the JWST, the EWs of strong \oiii\ and \ha\ emission lines at high redshifts were modeled from Spitzer/IRAC photometry. 
In this section, we compare the line EWs of galaxies in our sample, derived from rest-frame optical spectroscopy, with those inferred previously from the IRAC SED analyses.
}
% We first compare the line EWs of galaxies in our sample with those inferred previously from Spitzer/IRAC SED analysis.

In the left panel of Figure~\ref{fig:04_ew_ms}, we compare the \oiii$+$\hb\ line EWs and stellar masses with those of $z\sim6.8$ galaxies in \citet{endsley21b,endsley21a}.
The median \oiii$+$\hb\ line EW of the sources in our sample is \textred{416$\pm$66}\,\AA, with the maximum and minimum of \textred{1510$\pm$757}\,\AA\ (P330E-z6.15) and \textred{122$\pm$28}\,\AA\ (P330E-z6.28), respectively. 
Before the launch of JWST, \oiii$+$\hb\ line EWs at $z\gtrsim4$ could only be inferred from Spitzer/IRAC SED analysis in certain redshift windows, for example, $z\simeq 6.7 - 7.0$ \citep[e.g.,][]{smit14,smit15,endsley21b,endsley21a}.
\citet{endsley21a} reported a median \oiii$+$\hb\ EW of $759_{-113}^{+112}$\,\AA\ for a sample of UV-bright ($M_\mathrm{UV} \lesssim -21$) galaxies at $z\sim6.8$, and the scatter is $\sim$\,0.25\,dex.

Through a Kolmogorov–Smirnov test, we confirm that the \oiii$+$\hb\ line EW distribution of the sources in our sample is consistent with that in \citet{endsley21b,endsley21a} over a similar stellar-mass  ($\sim 10^9$\,\msun) and SFR ($\gtrsim 10$\,\smpy) range.
Similar to the findings in \citetalias{sun22}, this confirms the presence of strong rest-frame optical nebular emission lines in EoR galaxies with a wide range of line EWs ($\sim$1\,dex span).
\textred{Such a wide span is also seen with certain simulations \citep[e.g.][]{ceverino21}.}

We derive a median \ha\ EW of \textred{239$\pm$45}\,\AA\ for the galaxies in our sample with a maximum and minimum of \textred{814$\pm$489\,\AA\ (P330E-z6.15)} and \textred{84$\pm$29}\,\AA\ (P330E-z6.28), respectively.
We also compare the \ha\ EWs with those of $z\simeq5.1-5.4$ galaxies reported in \citet{rasappu16}, whose line EWs were inferred from Spitzer/IRAC SEDs.
In this redshift window, the \ha\ lines enter the passband of IRAC Channel\,2 (CH2; 4.5\,\micron), while the Channel\,1 (CH1, 3.6\,\micron) is free from strong emission lines including \hb\ and \oiii\ ($\lambda < 3.2$\,\micron).  Therefore, a red [3.6]--[4.5] color can be used to infer the strength of \ha\ emission.
We assumed that $\sim$\,80\%\ of the combined \ha+\nii+\sii\ EWs reported in \citet{rasappu16} is from \ha, similar to the fraction assumed in their work (84\%; from \citealt{af03}).

With comparable stellar masses and SFRs, the median \ha\ EW of the galaxies with spectroscopic redshifts in \citet{rasappu16} is $564\pm59$\,\AA, \textred{$2.4\pm0.5$}\,times that for our sample.
Although this can be potentially explained by the limited understanding of dust extinction and underlying stellar continuum in early Spitzer/IRAC studies, we cannot draw any firm conclusion from the \ha\ EW comparison given the small sample size.

\textred{
Finally, we also compare the line EWs with those of galaxies in the local Universe with comparable stellar masses.
We select galaxies at $z<0.05$ from the MPA-JHU value-added catalog of SDSS Data Release 7 \citep{kauffmann03b,sdssdr7} for comparison (black dots in Figure~\ref{fig:04_ew_ms}).
The EWs of \ha\ and \oiii\ lines of galaxies in our sample are higher than the median EWs of SDSS-selected galaxies by at least an order of magnitude. 
Our conclusion remains valid if we compared with galaxies in the Portsmouth SDSS catalog \citep{maraston13,thomas13}.
}

\subsection{Line Ratios and BPT Diagram}
\label{ss:04c_bpt}

% Among our sample, the \nii\,$\lambda$6583 lines of P330E-z6.35 was tentatively detected ($\sim1.5\sigma$), and a \nii/\ha\ ratio of 0.18$\pm$0.13 can be measured for both sources. 
% We note that P330E-z6.11 has two components with a physical offset of $\sim$\,1.6\,kpc and a velocity offset of $\sim$\,600\,km/s, and \nii/\ha\ ratio can be different for each component (see \citetalias{sun22}).
% In this work, we adopt the global line ratio for the analysis.
% P330E-z6.11 is also the only source with a tentative \hb\ detection ($1.5\sigma$), and the \oiii\,$\lambda$5007/\hb\ line ratio is 8$\pm$5.
% For other sources, the typical $3\sigma$ lower limit of this line ratio is $\sim$2.

\textred{Despite secure identification of \oiii\ and \ha\ lines, the sources in our sample are typically undetected in either \nii\,$\lambda$6583 or \hb\ lines. 
With the latest calibrated spectra, the \hb\ line of P330E-z6.11 is detected at 2.5$\sigma$, suggesting a line ratio of \oiii/\hb\,$=4.9\pm1.3$. 
None of the sources are detected in \nii\,$\lambda$6583 at above $2\sigma$, and thus the typical $3\sigma$ upper limit of \nii/\ha\ line ratio is $<0.4$.
}

With these line ratios and upper/lower limits, we plot our sources on the \oiii/\hb-\nii/\ha\ BPT diagram \citep{bpt1981} in Figure~\ref{fig:05_bpt}.
We compare our sample with galaxies in the local Universe \textred{\citep[in the MPA-JHU catalog for SDSS Data Release 7;][]{sdssdr7}} and $z\sim2.3$ galaxies in the MOSDEF sample \citep{kriek15,reddy15,shapley15}.
Although the detection rate of \hb\ and \nii\ lines is low, the $z>6$ emission-line galaxies in our sample appear to occupy the same parameter space as that of $z\sim2$ star-forming galaxies in the MOSDEF sample, and are likely located above the star-forming sequence of SDSS galaxies \textred{(see also recent studies with NIRSpec, e.g., \citealt{cameron23} and \citealt{sanders23}).}
Similar to $z\sim2$ galaxies, this could be explained by an elevated N/O abundance at a given O/H ratio and/or a higher ionization parameter \citep[e.g.,][]{kewley13,masters14,steidel14,shapley15,kojima17,curti22a} for high-redshift galaxies with moderate stellar masses ($<10^{10}$\,\msun).
\textred{Stars that are enhanced in alpha elements can have harder intrinsic ionization spectra, which can lead to a higher \oiii/\hb\ at fixed \nii/\ha\ in nebulae \citep[e.g.,][]{strom17,topping20a,topping20b}.
}

\textred{
It is possible that sources in our sample may contain active galactic nuclei (AGN).
However, based on their locations on the BPT diagram \citep{kewley01, kauffmann03}, hydrogen line profiles and number density (Section~\ref{sec:05_vol}), we do not find strong evidence to confirm any source as AGN.
}

\subsection{Gas-Phase Metallicity}
\label{ss:04d_metal}

\textred{With oxygen and hydrogen lines detected for all sources in our sample}, we study their gas-phase metallicity using the strong-line calibrations of \citet{bian18}.
\citet{bian18} studied the stacked spectra of local analogs of $z\sim2$ star-forming galaxies, derived direct gas-phase O/H abundances using \oiii\,$\lambda$4363 lines, and established empirical metallicity calibrations between O/H abundances and strong-line ratios such as \nii\,$\lambda$6583/\ha\ (also known as N2), \oiii\,$\lambda$5007/\hb, (\oiii\,$\lambda\lambda$4959,5007$+$\oii\,$\lambda$3727)/\hb\ (also known as R23) and \oiii\,$\lambda\lambda$4959,5007/\oii\,$\lambda$3727 (also known as O32).
Given the similarity of the locations in the \oiii/\hb--\nii/\ha\ BPT diagram (Figure~\ref{fig:05_bpt}) among our sample, $z\sim2$ star-forming galaxies and the local analogs analyzed in \citet{bian18}, these empirical calibrations are likely useful for these galaxies at $z>6$.
In the absence of statistical samples of strong line ratios above $z>3$, an estimate using the $z=0-2$ calibrations can provide a valuable first look, but we also make the caveat clear that the empirical calibration may break down in this unexplored redshift regime (e.g., see discussions in \citealt{curti22} \textred{and most recently \citealt{sanders23b}}).
This is possibly because of the different ISM conditions and physical properties (e.g. N/O, ionization parameter, hardness of the ionizing radiation; \citealt{steidel16}, \citealt{strom17}) of high-redshift galaxies relative to the calibration samples at lower redshifts \textred{\citep{brinchmann22}}.
% the uncertainties from N/O abundance and ionizing radiation field \citep[e.g.,][]{steidel16,strom17}.

\textred{We derive the gas-phase metallicity of each source by averaging the measurements with multiple tracers,
including N2 (P330E-z6.35), R23 (P330E-z6.28 and P330E-z6.35) and \oiii/\hb\ (O3; all sources).
For sources without \hb\ detections, we assume \textred{an intrinsic} \ha/\hb\ ratio of 2.86 (case-B recombination and electron temperature of $10^4$\,K)}.
One caveat is that the \ha/\hb\ ratio could be underestimated because of the dust extinction, and the resultant O/H abundance could be overestimated by 0.2\,dex if the \ha/\hb\ ratio is 4.
% For P330E-z6.28, O32 and \oiii/\hb\ methods suggest a $12+\log(\mathrm{O/H})$ value of $8.5\pm0.2$.
\textred{We estimated the uncertainty of metallicity from both the errors of line ratios and the scattering of measurements with multiple tracers.
The derived gas-phase metallicities are reported in Table~\ref{tab:01_prop}.}
In general, the galaxies in our sample have been enriched to moderate metallicities (\textred{$\sim$0.4\,\zsun}).

Figure~\ref{fig:06_mzr} shows the metallicities versus stellar masses of sources in our sample at $z>6$.
We also compare our measurements with those of the four $z>6$ galaxies observed through the JWST/NIRSpec ERO of the SMACS0723 field \citep{pontoppidan22}.
For three sources at $z>7$, we adopt the stellar mass measurements from \citet[see also \citealt{trussler22}]{tacchella22} and gas-phase metallicities from \citet{curti22} using the direct $T_\mathrm{e}$ method.  
For the $z=6.38$ galaxy, we adopt the $M_\mathrm{star}$ from \citet{carnall22} and $12+\log(\mathrm{O/H})$ from \citet{taylor22}.  
We also note that metallicity measurements from \citet{rhoads22}, \citet{schaerer22}, \citet{tacchella22} and \citet{trump22} are in general agreement \citep{brinchmann22}.
With higher stellar masses and slightly lower redshift, galaxies in our sample exhibit higher metallicities than those of the four $z>6$ galaxies in the SMACS0723 field.
The best-fit mass-metallicity (gas-phase) relation from these eight sources at $z=6.0 - 8.5$ is:
\begin{equation}
    12 + \log(\mathrm{O/H}) = (0.57\pm0.16) \log(\frac{M_\mathrm{star}}{10^9\,\msun}) + 8.18\pm0.12
\end{equation}
which is shown as the solid orange line in Figure~\ref{fig:06_mzr} with the lighter filled region indicating the $1\sigma$ uncertainty range.

Within the investigated stellar mass range ($M_\mathrm{star}=10^{7.5} \sim 10^{9.5} $\,\msun), our best-fit relation suggests a lower metallicity at a given stellar mass when compared with $z\sim0$ galaxies observed with SDSS (e.g., \citealt{curti20}).
However, four galaxies in our sample exhibit moderate metallicities that are comparable to those of $z\simeq2 - 3$ galaxies in the MOSDEF sample \citep{sanders21}, whose metallicities were also derived based on the \citet{bian18} calibration.
The observed metallicities of sources in our sample are \textred{slightly} higher than those of galaxies with similar stellar masses in certain cosmological simulations including IllustrisTNG \citep{torrey19}.
As a result, the best-fit mass-metallicity relation at $z>6$ with JWST is also steeper than that in IllustrisTNG.

However, we note that the slope of the mass-metallicity relation flattens to $0.41\pm0.07$ if the $z=8.5$ source with extremely low metallicity ($12+\log(\mathrm{O/H})=7.0\pm0.1$; \citealt{curti22}) in JWST/NIRSpec ERO is excluded from the linear fitting, making the slope closer to those seen at $z\simeq2-3$ ($\sim0.30$, \citealt{sanders21}; \textred{see also \citealt{limy23} most recently}).
We also note that to fully explore the evolution with respect to low-redshift galaxies and compare with simulations, one should also consider the secondary dependence of metallicity on SFR (i.e., the fundamental metallicity relation, see discussion in \citealt{curti22}).
This additional step is beyond the scope of this paper.
% We conclude that the O/H abundance of $z>6$ galaxies in our sample and resultant steep slope of the mass-metallicity relation 

Our observations may indicate a rapid metal enrichment in certain massive ($M_\mathrm{star}\gtrsim10^9$\,\msun) galaxies at $z>6$, which is also suggested by the \oiii\,88\,\micron\ and dust detections of $z>8$ galaxies with ALMA (\citealt{tamura19}; see also \citealt{jones20}).
The enhanced gas-phase metallicity potentially indicates that the recent increase in SFR is driven by mergers or internal gravitational instabilities \citep[e.g.,][]{tacchella22} instead of pristine gas inflow, which can result in lower gas-phase metallicity.
Such ancient and rapid metallicity evolution and intense episode of star formation is also in a sense reminiscent of the formation scenario of our own Milky Way Bulge, which indeed rapidly evolved toward solar metallicity and formed most of stellar mass $>10$\,Gyr ago.
However, we also note that our metallicity measurements are totally based on the low-redshift strong-line calibrations, mostly in the metallicity range of $7.8 < 12 + \log(\mathrm{O/H}) < 8.4$ \citep{bian18}.
Further direct gas-phase metallicity measurements through JWST/NIRSpec \textred{observations (e.g., most recently with \citealt{nakajima23} and \citealt{sanders23b})}
%(such as those in the SMACS0723 field; \citealt{pontoppidan22})} 
are necessary for more accurate determination of the mass-metallicity relation in the high-redshift Universe.

\subsection{Redshift Evolution of \ha\ line Equivalent Width}
\label{ss:04e_ha}

\begin{figure*}
\centering
\includegraphics[width=0.49\linewidth]{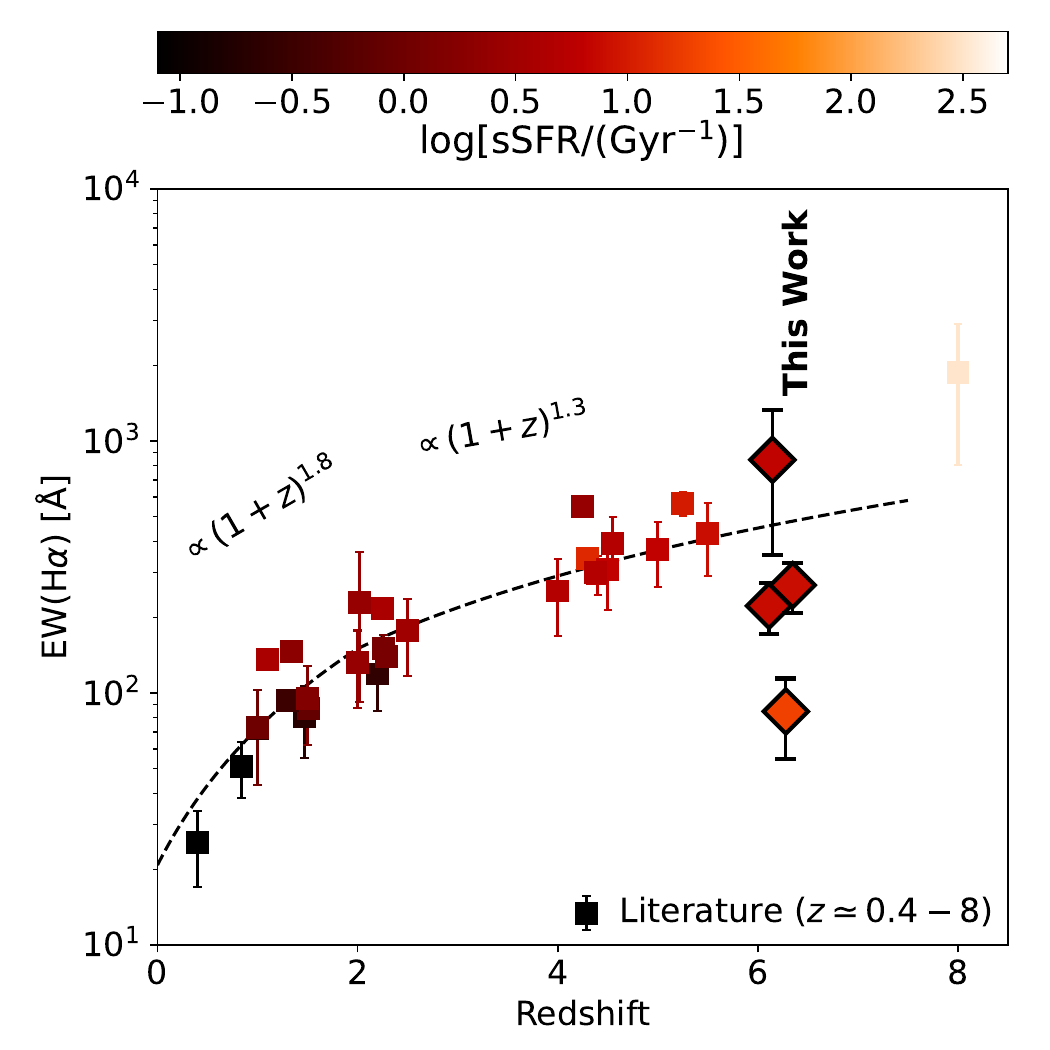}
\includegraphics[width=0.49\linewidth]{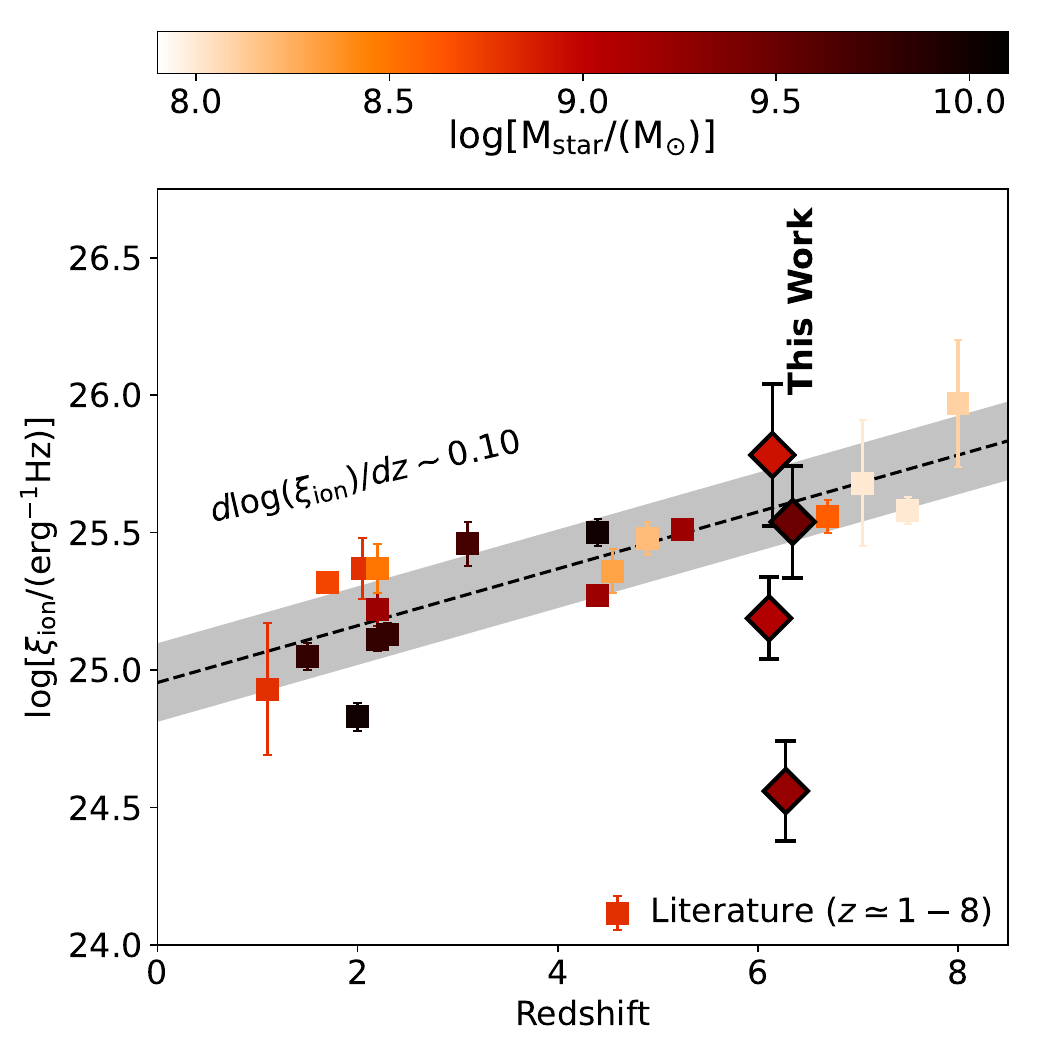}
\caption{Redshift evolution of \ha\ equivalent widths (left) and ionizing photon production efficiency (right) of star-forming galaxies across $z \simeq 0.4 - 8$.
Sources in our sample are shown as diamonds.
In the left panel, the reference samples (squares) include \citet{erb06}, \citet{shim11}, \citet{fumagalli12}, \citet{stark13}, \citet{sobral14}, \citet{faisst16},  \citet{mq16}, \citet{rasappu16}, \citet{smit16}, \citet{reddy18}, \citet{lam19}, \citet{atek22}, \citet{boyett22} and \citet{stefanon22}, and all samples are color-coded by their specific SFRs.
The proposed redshift evolution ($\mathrm{EW}\propto (1+z)^{1.8}$ at $z<2$ and $\propto(1+z)^{1.3}$ beyond) in \citet{faisst16} is shown as the dashed black line.
In the right panel, the reference samples (squares) include \citet{stark15,stark17}, \citet{bouwens16a}, \citet{nakajima16}, \citet{matthee17},  \citet{harikane18},  \citet{shivaei18},  \citet{faisst19}, \citet{lam19},  \citet{tang19},  \citet{emami20}, \citet{nanayakkara20},  \citet{endsley21b}, \citet{atek22} and \citet{stefanon22}, and all samples are color-coded by their stellar masses.
The best-fit redshift evolution ($d\log\xi_\mathrm{ion}/dz =0.10\pm0.02$) of literature samples is shown as the dashed black line, and the $1\sigma$ dispersion of the relation (0.14\,dex) is indicated by the shaded gray region. 
}
\label{fig:07_evolution}
\end{figure*}

\textred{Before the launch of the JWST, the \ha\ EWs at $z\sim6$ were poorly probed because \oiii\ and \ha\ lines are in the bandwidth of IRAC Channel 1 and 2, respectively.}
In the left panel of Figure~\ref{fig:07_evolution}, we examine the redshift evolution of \ha\ EWs of galaxies from $z\simeq0.5$ to 8 in the literature.
For \ha\ EW measurements obtained with medium/high-resolution spectroscopy \citep[e.g.,][]{erb06,reddy18}, we directly use the reported EWs.
For measurements with low-resolution grism spectroscopy \citep[e.g.,][]{fumagalli12,atek22,boyett22} or narrow-band imaging \citep{sobral14} in which \ha\ lines are blended with \nii\ lines, we assume that 85\% of the line fluxes are from \ha, consistent with our \nii/\ha\ measurements in Section~\ref{ss:04c_bpt} \textred{(see also recent JWST studies, e.g., \citealt{cameron23}, \citealt{helton23} and \citealt{sanders23})}.
For measurements relying on broad-band photometry and SED modeling \citep[e.g.,][]{shim11,stark13,mq16,rasappu16,smit16,lam19,stefanon22}, we assume that 80\% of \ha$+$\nii$+$\sii\ line fluxes are from \ha, consistent with the fractions adopted by various studies (71--84\%; e.g., \citealt{shim11}, \citealt{rasappu16}).

The redshift evolution of \ha\ EWs has been described with $\mathrm{EW} \propto (1+z)^{1.8}$ at $z\lesssim 2$ (e.g., \citealt{fumagalli12}) and $\propto (1 + z)^{1.3}$ at $z > 2$ \citep{faisst16}.
At $z>2.5$, all \ha\ EWs published before the JWST were derived based on Spitzer/IRAC broad-band colors, and therefore are inevitably affected by multiple uncertainties, including the assumption of SFH and dust attenuation.
The median EW(\ha) of sources in our sample is indeed higher than those of $z\lesssim 2$ galaxies with direct spectroscopic measurements and similar to the previous estimates at $z\simeq 4-5$ \citep[e.g.,][]{shim11,stark13,faisst16,mq16,rasappu16,smit16,lam19}.
This is likely a fair comparison because the specific SFRs (sSFR; SFR per unit $M_\mathrm{star}$) of these samples are comparable.  
Note that the \ha\ EWs broadly reflect the sSFR's of galaxies because \ha\ is a SFR tracer while the underlying continuum is \textred{related to the luminous} stellar mass.
However, as the redshifts of galaxies in our sample are higher than those measured with Spitzer/IRAC [3.6]--[4.5] color, this may indicate that the redshift evolution of the \ha\ EWs is flattening toward the EoR, although the robustness of our conclusion is limited by the small sample size. 
Finally, our \ha\ EW measurements are much lower than that of stacked $z\sim8$ galaxies ($\sim2\times10^3$\,\AA) as inferred from Spitzer/IRAC [3.6]--[5.8] color \citep{stefanon22} despite a large uncertainty.

\subsection{Redshift Evolution of $\xi_\mathrm{ion}$}
\label{ss:04f_xi}

Following a few studies including \citet{matthee17}, \citet{shivaei18}, \citet{tang19} \textred{and many others}, we derive the ionizing photon production efficiency in \hii\ regions as $\xi_\mathrm{ion} = N(\mathrm{H}^0) / L_\mathrm{UV}$, where $N(\mathrm{H}^0)$ is the ionizing photon production rate in the unit of \si{s^{-1}}, and $L_\mathrm{UV}$ is the rest-frame UV luminosity at 1500\,\AA\ in the unit of \si{erg.s^{-1}.Hz^{-1}}.
The ionizing photon production rate can be computed from the \ha\ luminosity as $N(\mathrm{H}^0) = 7.35\times10^{11} L_\mathrm{H\alpha} $ where $L_\mathrm{H\alpha}$ is in the unit of \si{erg.s^{-1}} (\citealt{osterbrock06}; Case B recombination at $T_\mathrm{e}=10^4$\,K).
This is true assuming that the Lyman continuum escape fraction is low ($\lesssim10$\%).
Note, however, that we do not have near-infrared (1.0--2.2\,\micron) photometry for our galaxies sampling the rest-frame UV continuum, and therefore we estimate the rest-frame UV luminosities based on the \textsc{cigale} modeling performed in the rest-frame optical as already described.

The median derived $\log[\xi_\mathrm{ion}/(\mathrm{erg}^{-1}\,\mathrm{Hz})]$ of galaxies in our sample is $25.4\pm0.2$, higher than the canonical value of $\sim 25.1$ implied by the UV--SFR and \ha--SFR conversion in \citet{ke12}.
However, a stronger dust attenuation in the rest-frame UV could lead to an overestimate of $\xi_\mathrm{ion}$.
Assuming \citet{calzetti00} extinction law, with an $A_V=1.0$ and a canonical $E(B-V)$ ratio of 0.44 between stellar continuum and nebular lines, $\xi_\mathrm{ion}$ would be overestimated by 1.9\,times if the dust attenuation is not properly corrected (see also \citealt{shivaei18}).
% {\color{red} (Remove the following line.  It is too general and obvious.)}
% \st{Although the accuracy of dust attenuation modeling in this work is limited by the shallowness of commissioning data, further JWST/NIRCam grism observations in cosmological deep field with more filters will provide better measurements of the ionizing production efficiency at $z\gtrsim 6$.}

In the right panel of Figure~\ref{fig:07_evolution}, we study the redshift evolution of the ionization photon production efficiency.
The published samples at $z\simeq 1 - 8$ include \citet{stark15,stark17}, \citet{bouwens16a}, \citet{nakajima16}, \citet{matthee17},  \citet{harikane18},  \citet{shivaei18},  \citet{faisst19}, \citet{lam19},  \citet{tang19},  \citet{emami20}, \citet{nanayakkara20},  \citet{endsley21b}, \citet{atek22} and \citet{stefanon22}. 
The stellar mass range of the reference sample is $M_\mathrm{star}\simeq 10^{8} - 10^{10}$\,\msun, comparable to that of our sample except for those at very high redshifts ($z>7$).
Despite a considerable uncertainty of UV luminosity (because we rely on best-fit SED models) and dust attenuation, the $\xi_\mathrm{ion}$ measured in this work based on accurate \ha\ luminosity is consistent with the previous determination at $z\simeq 5 - 7$  (e.g., \citealt{stark15}, \citealt{harikane18}, \citealt{endsley21b} \textred{and most recently \citealt{ning23} and \citealt{tang23}}).

As shown by the dashed black line in the plot, the published samples indicate a redshift evolution of $\xi_\mathrm{ion}$ with a slope of $d\log(\xi_\mathrm{ion})/dz = 0.10\pm0.02$ (see also \citealt{matthee17}, \citealt{atek22} and \citealt{stefanon22}).
This redshift evolution can be interpreted by an age effect, i.e., galaxies at higher redshifts have younger stellar populations therefore higher $\xi_\mathrm{ion}$ \citep[e.g.,][]{tacchella18,naidu20}.
The galaxies in our sample are in general agreement with such a redshift evolution trend except for P330E-z6.28 ($\log[\xi_\mathrm{ion}/(\mathrm{erg}^{-1}\mathrm{Hz})] = 24.6 \pm 0.2$).
The large scatter of $\log(\xi_\mathrm{ion})$ in our sample (\textred{0.46}\,dex; including P330E-z6.28) is also not a surprise because it is also seen in other samples at lower redshifts, which can be propagated from the scatter of dust attenuation, sSFR and patchy ISM coverage (e.g., \citealt{matthee17}, \citealt{shivaei18}).

\section{Discussion \RomanNumeralCaps{2}: Volume Density}
\label{sec:05_vol}

\subsection{Emission Line Luminosity Function: Methodology}
\label{ss:05a_method}

To compute the LFs of \oiii\,$\lambda$5007 and \ha\ lines at $z\sim 6.2$, we first compute the total survey volume of \oiii$+$\ha\ emitters at $z=6.0\sim6.6$.
With the spectral tracing model, dispersion model and pointing information of the telescope, we compute the maximum survey area in which the direct-imaging sources at $z>6$ could yield both detectable \oiii\ and \ha\ emission lines with the obtained F322W2 and F444W grism observations.
The resultant survey area of \oiii$+$\ha\ emitters at $z=6.2$ is shown as the purple contours in Figure~\ref{fig:01_map}.
The maximum survey area changes from 14.4\,\si{arcmin^2} at $z=6.0$ to 12.4\,\si{arcmin^2} at $z=6.6$ because the overlapping area that could yield dual line detections decreases toward higher redshifts.

With the maximum survey area, we compute the emission-line LF without a completeness correction, which we refer to as the \textit{uncorrected} LF ($\Phi_\mathrm{uncorr}$).
This reflects the lower limits of \ha\ and \oiii\ LF measurements at $z\sim6.2$ that are directly inferred from the number of source detections ($N_\mathrm{src}$), maximum survey volume ($V_\mathrm{max}$ in comoving \si{Mpc^3}; assuming $z=6.0-6.6$) and luminosity bin size ($d\log L$ in the unit of dex) as $\Phi_\mathrm{uncorr} = N_\mathrm{src} / (V_\mathrm{max}\, d\log L) $.
The uncorrected LFs of \ha\ and \oiii\,$\lambda$5007 lines are computed in luminosity bins of \textred{$10^{42.4}-10^{42.8}$ and $10^{42.8}-10^{43.2}$\,\si{erg.s^{-1}}}, and Monte Carlo (MC) simulations are performed to quantify the uncertainty propagated from the line flux errors.
We also consider the Poisson noise of small number statistics using the prescription of \citet{gehrels86}.
The uncorrected LFs are reported in Table~\ref{tab:02_lf}.

We also compute the line LFs using the direct $1/V_\mathrm{max}$ method \citep{schmidt68} as:
\begin{equation}
\Phi(L) = \frac{1}{d\log L} \sum_{i} \frac{1}{C_{i} V_{\mathrm{max},i}}
\end{equation}
where $C_{i}$ is the completeness of the $i$-th source in the luminosity bin, and $V_{\mathrm{max},i}$ is the maximum observable volume of the $i$-th source.

The survey completeness is evaluated through MC, which is detailed in Appendix~\ref{apd:01}.
In short, we compute the completeness of all line detections by injecting mock line emissions in the 2D spectral image, deriving the line fluxes and errors using the same method as we applied for the real line detections, and evaluating the fraction of realizations above the $3\sigma$ detection threshold.
We also evaluate the flux-boosting effect and the Eddington bias by comparing injected and output line flux ratios, and correct this for the line luminosities of sources in our sample, which could be boosted by 18\%\ for a S/N\,$=$\,3 line detection.
We also compute the maximum survey volume of each source from the RMS error map of the F322W2 and F444W grism images, in which the \oiii\,$\lambda$5007 and \ha\ lines with the same luminosities as those of the real source can be detected at $\geq 3\sigma$.
In general, the maximum survey area decreases from $z=6.0$ to 6.6 because of the increasing RMS noise toward the red end of the F444W filter. 
% The $z=6.35$ source has the largest maximum survey volume

We then compute the completeness-corrected line LFs ($\Phi_\mathrm{corr}$) using the same luminosity bins as those for $\Phi_\mathrm{uncorr}$.
We also employ MC simulations for correct error propagation from line flux to LF and consider Poisson noise for small-number statistics as described above for $\Phi_\mathrm{uncorr}$.
\textred{Finally, we also consider the impact from the cosmic variance following the simple prescription in \citet{driver10}, which could introduce a further uncertainty of 0.2\,dex.
% Because galaxies in our sample are confirmed at different redshifts, it is unlikely that our luminosity function measurements are affected by galaxy overdensities, which cluster at specific redshifts.
}

\begin{figure*}[!pt]
\centering
\includegraphics[width=0.49\linewidth]{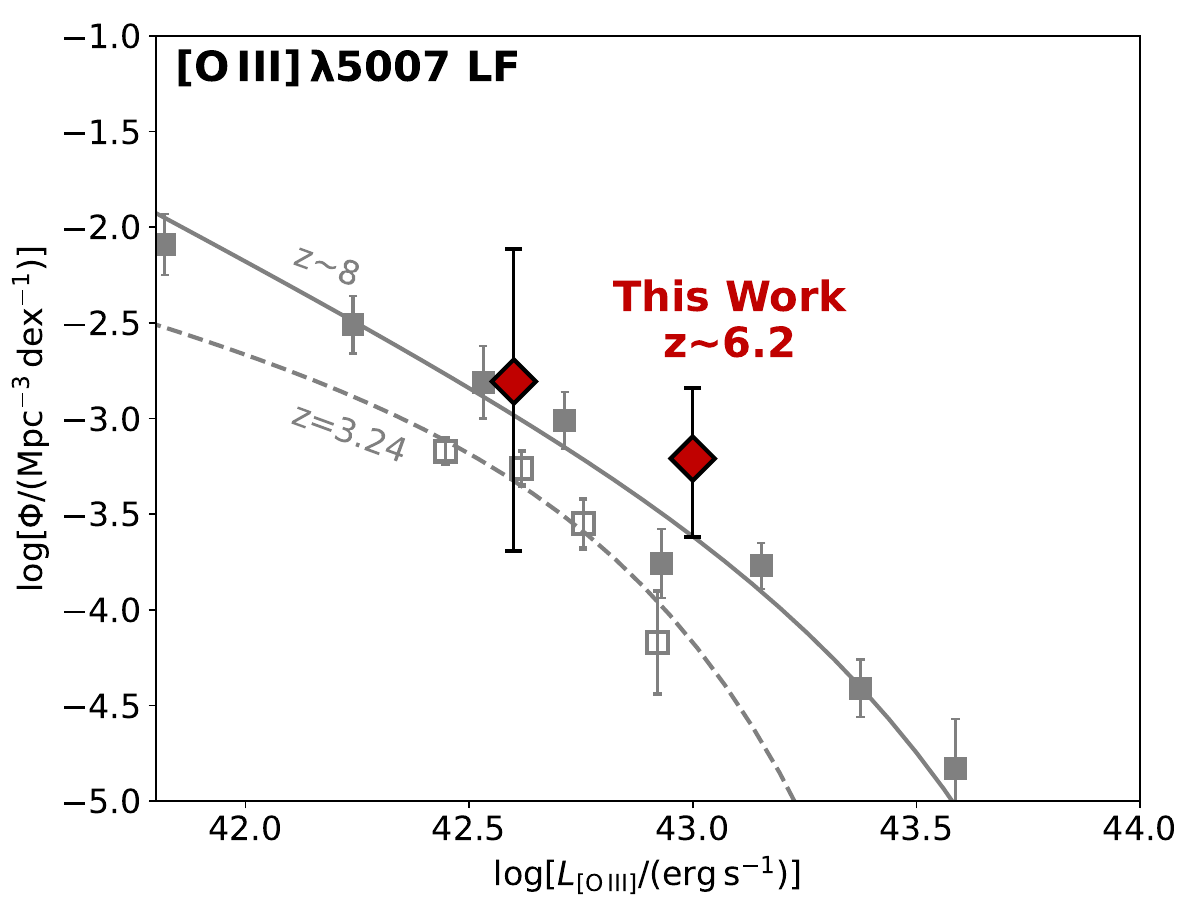}
\includegraphics[width=0.49\linewidth]{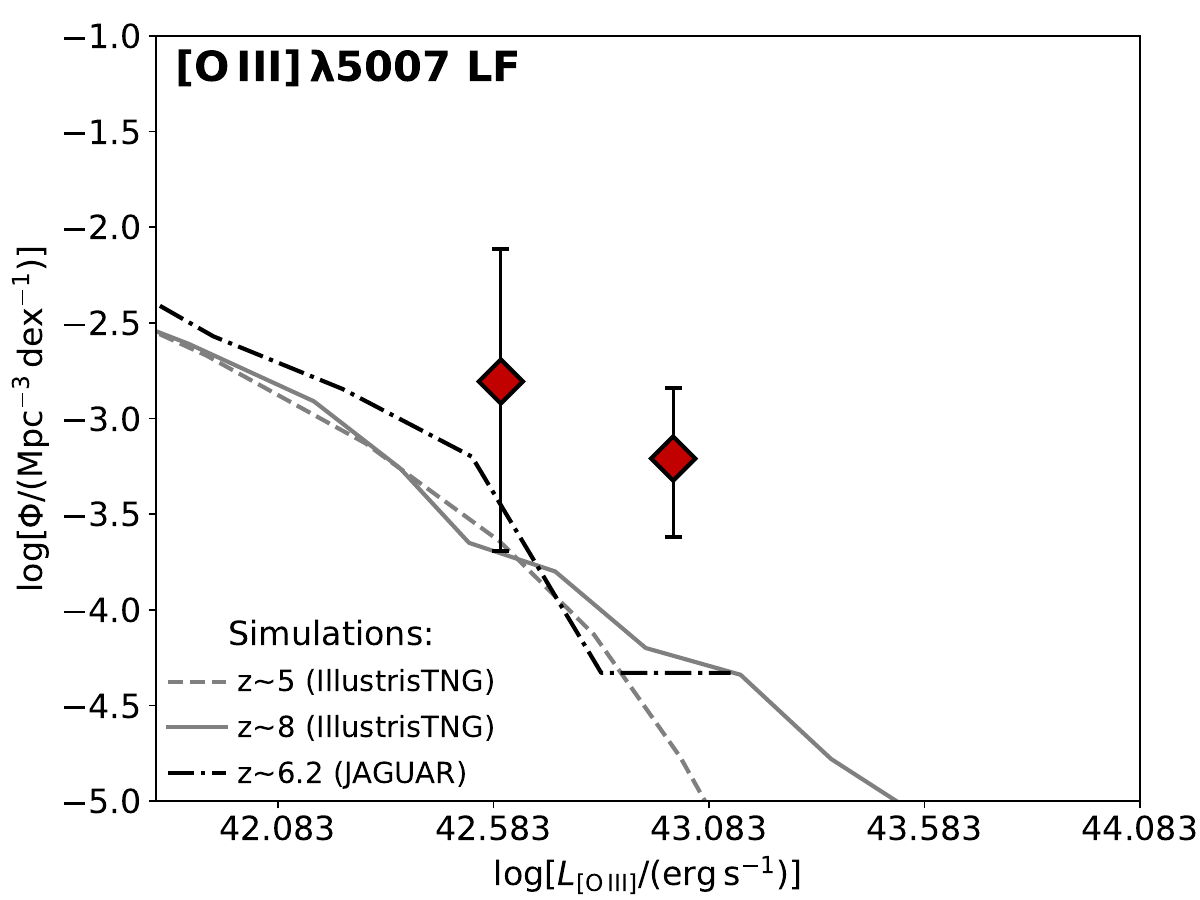}
\includegraphics[width=0.49\linewidth]{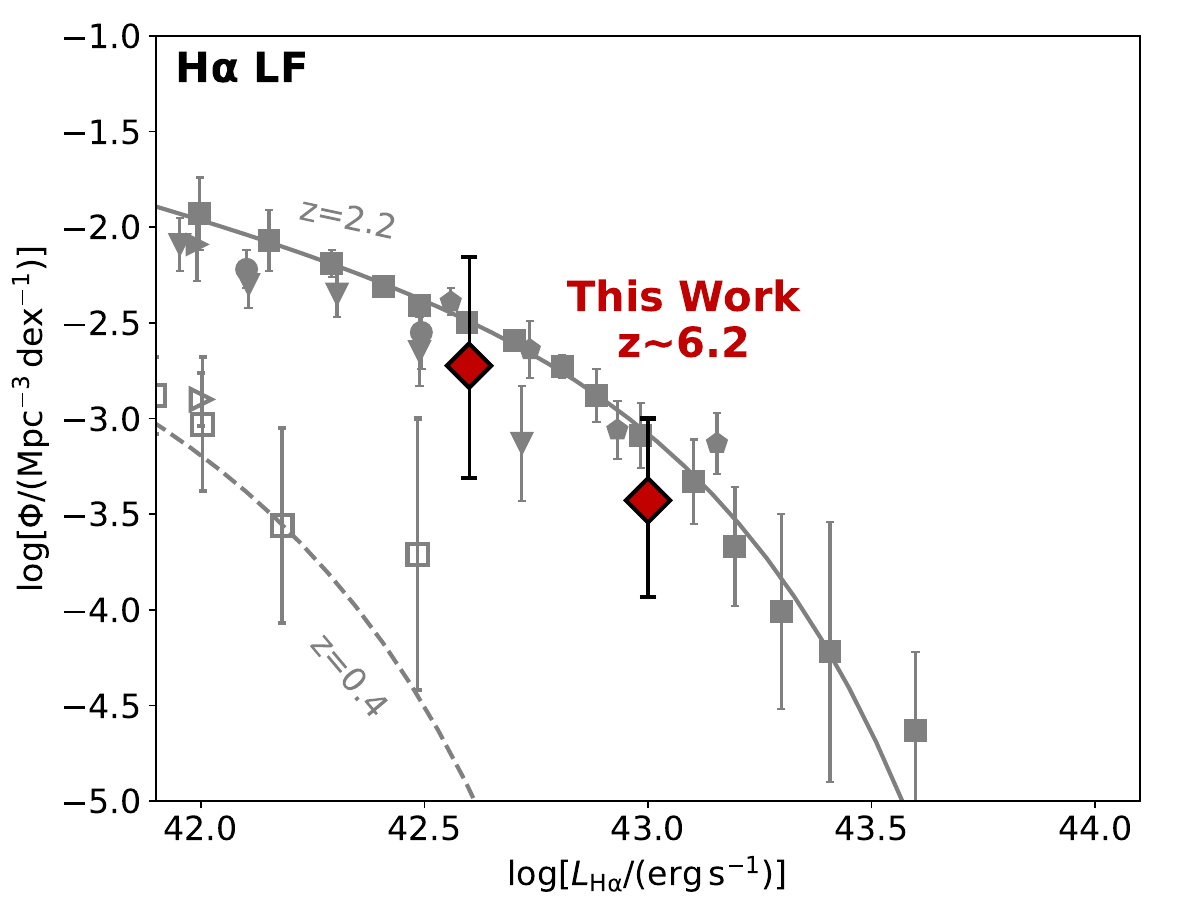}
\includegraphics[width=0.49\linewidth]{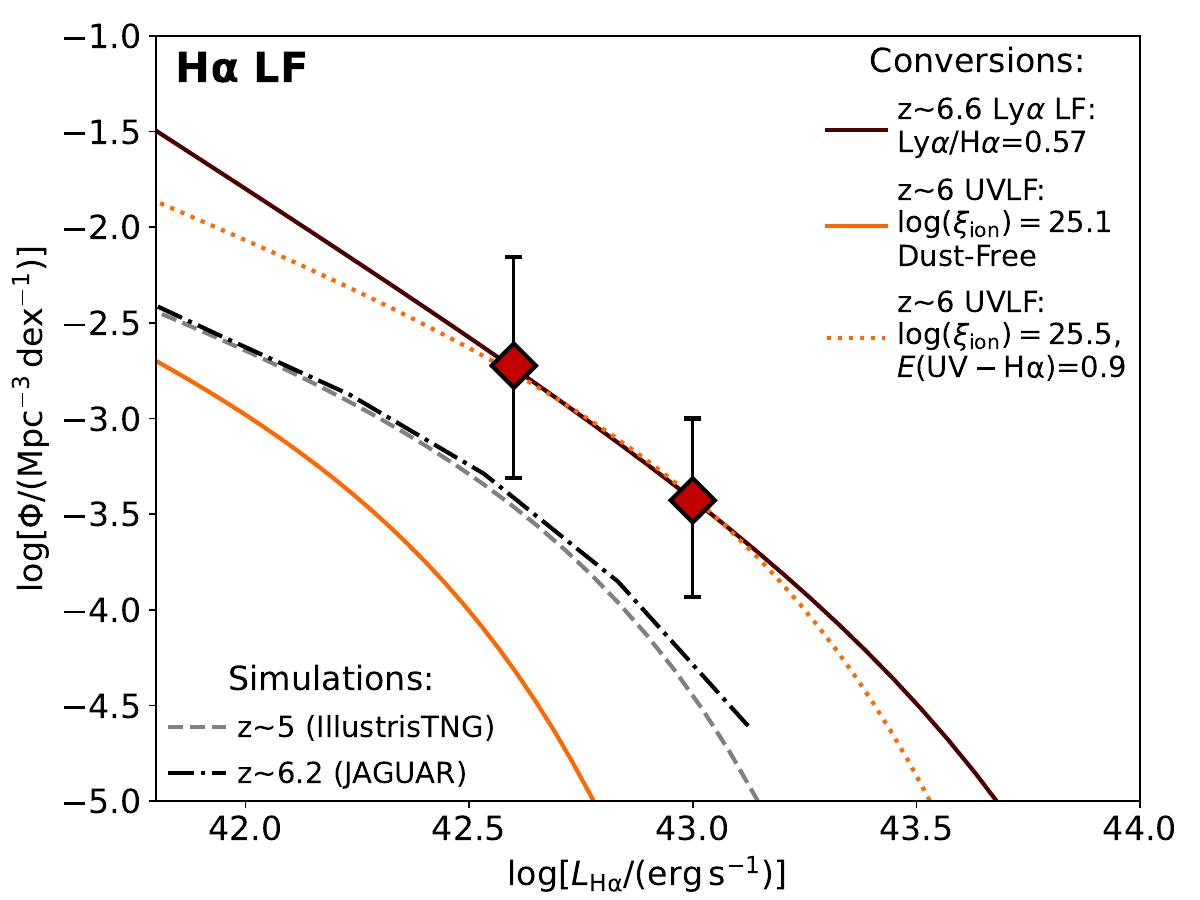}
\caption{\textbf{Top-left}: $z\sim6.2$ \oiii\,$\lambda$5007 line luminosity function derived in this work (red diamonds). 
For comparison we show the $z=3.24$ \oiii\ LF measured by \citet{khostovan15} through narrow-band imaging (open squares and dashed line), and $z\sim8$ \oiii\ LF \citep{debarros19} inferred from the UVLF and Spitzer/IRAC colors (filled squares and solid line).
\textbf{Top-right}: Observed \oiii\,$\lambda$5007 luminosity function compared with those in simulations, including IllustrisTNG at $z=5$ (dashed gray line), $z=8$ (solid gray line; \citealt{shen20}) and the JAGUAR mock catalog at $z\sim6.2$ (dash-dotted black line; \citealt{williams18}).
\textbf{Bottom-left}: $z\sim6.2$ $\ha$ line luminosity function derived in this work (red diamonds).
For comparison, we show the \ha\ LF measurements at $z=2.2$ (\citealt{geach08}, \citealt{hayes10}, \citealt{tadaki11}, \citealt{lee12}, \citealt{sobral13}) and $z=0.4$ (\citealt{ly07}, \citealt{sobral13}).
Best-fit Schechter functions at $z=2.2$ and 0.4 \citep{sobral13} are shown in solid and dashed gray lines, respectively.
\textbf{Bottom-right}: observed \ha\ LF compared with those in simulations (IllustrisTNG at $z\sim5$, dashed gray lines, \citealt{shen20}; JAGUAR mock catalog at $z \sim 6.2$, dash-dotted black line, \citealt{williams18}).
We also compare with the $z\sim6$ UVLF \citep{bouwens21a} by converting UV luminosity to \ha\ luminosity under a constant SFR assumption (\citealt{ke12}; effectively $\log[\xi_\mathrm{ion}/(\mathrm{erg}^{-1}\mathrm{Hz})]=25.1$ and no dust attenuation) as shown in solid orange line.
\textred{We also perform the conversion assuming} an enhanced $\log[\xi_\mathrm{ion}/(\mathrm{erg}^{-1}\mathrm{Hz})]=25.5$ and $\mathrm{UV}-\ha$ color excess of \textred{0.9} as shown in dotted orange line.
Conversion from $z\sim6.6$ \lya\ LF (\citealt{konno18}; assuming a line ratio of \lya/\ha\,$=$\,\textred{0.57}) is shown in solid brown line.
}
\label{fig:08_lf}
\end{figure*}

The corrected LFs are also reported in Table~\ref{tab:02_lf}.
In the bright bin ($L_\mathrm{line} = 10^{42.8} - 10^{43.2}$\,\si{erg.s^{-1}}), $\log(\Phi_\mathrm{corr})$ is higher than the uncorrected value by \textred{0.1--0.2\,dex}.
In the faint bin ($L_\mathrm{line} = 10^{42.4} - 10^{42.8}$\,\si{erg.s^{-1}}), the uncorrected LF likely underpredicts the volume density of both \ha\ and \oiii\,$\lambda$5007 emitters by $\sim 1$\,dex, although with a large uncertainty ($0.6\sim0.9$\,dex) because of small number statistics and a large uncertainty propagated from the line flux error.
In Figure~\ref{fig:08_lf}, we display the measured $\log(\Phi_\mathrm{corr})$ of \oiii\,$\lambda$5007 and \ha\ line-emitters at $z\sim 6.2$, and compared them with other observations and simulations.
We note that this is the first time that one could directly measure the LFs of both lines in the EoR, thanks to the unprecedented sensitivity of JWST beyond the $K$ band and unique ability of the NIRCam/Grism WFSS mode to sample line emitters in an unbiased way.
However, we are still not able to model the LFs with commonly used formalisms (e.g., Schechter function; \citealt{schechter76}) because of the limited sample size.

%% line LF table
\begin{table}[!t]
\caption{Measured \ha\ and \oiii\,$\lambda$5007 line luminosity function at $z\sim6.2$.}
% \centering
\begin{tabular}{cccc}
\hline\hline
$\log (L_\mathrm{line})$ & $N_\mathrm{src}$ & $\log(\Phi_\mathrm{uncorr})$ & $\log(\Phi_\mathrm{corr})$ \\
\textred{(\si{erg.s^{-1}})}  &  & (\si{Mpc^{-3}.dex^{-1}}) & (\si{Mpc^{-3}.dex^{-1}}) \\
\hline\multicolumn3l{\oiii\,$\lambda$5007} \\\cline{0-1}
42.6 & 1 & $-4.07_{-0.76}^{+0.64}$ & $-2.81_{-0.89}^{+0.69}$  \\
43.0 & 3 & $-3.36_{-0.32}^{+0.28}$  & $-3.21_{-0.41}^{+0.37}$   \\
\hline\multicolumn3l{\ha} \\\cline{0-1}
42.6 & 2 &  $-3.64_{-0.47}^{+0.40}$    & $-2.72_{-0.59}^{+0.57}$ \\
43.0 & 2 & $-3.54_{-0.41}^{+0.35}$   & $-3.43_{-0.50}^{+0.43}$  \\\hline
\end{tabular}
\tablecomments{Columns: (1) $\log (L_\mathrm{line})$: Center of emission-line luminosity bin (bin size: 0.4\,dex) in unit of \textred{\si{erg.s^{-1}}}; (2) $N_\mathrm{src}$: Number of sources in the luminosity bin; (3)/(4): uncorrected and corrected volume density of line emitters within the given luminosity bin, respectively (see Section~\ref{ss:05a_method}).
}
\label{tab:02_lf}
\end{table}

\subsection{\oiii\,$\lambda$5007 Luminosity Function}
\label{ss:05b_o3lf}

The \oiii\,$\lambda$5007 LF at $z\sim6.2$ is shown in the top-left panel of Figure~\ref{fig:08_lf}.
For comparison, we also plot the \oiii\,$\lambda$5007 LF measured by \citet{khostovan15} at $z=3.24$, %through the 2.12\,\micron\ narrow-band filter,
the highest-redshift \oiii\ LF measurement that is accessible from the ground using the rest-frame H$_{2}$ 1--0 S(1) narrow-band filter in the $K$ band (2.12\,\micron).
We also compare our results with the $z\sim8$ \oiii\ LF derived by \citet{debarros19} based on the UV LF and the relation between UV and \oiii+\hb\ luminosities as inferred from Spitzer/IRAC [3.6]--[4.5] colors.
For a fair comparison, we assumed that the  \oiii\,$\lambda$5007 line contributes to $\sim64$\% of the total fluxes from \hb+\oiii\ lines in the LFs of \citet{khostovan15} and \citet{debarros19}.
The adopted value is consistent with the median fraction observed with our sample, which is computed as $f_\mathrm{[O\,III]\,5007}/(1.33\, f_\mathrm{[O\,III]\,5007} + f_\mathrm{H\alpha} / 2.86)$, assuming the theoretical ratio of 1/3 for \oiii\,4959/5007 lines and 2.86 for \ha/\hb\ lines.

We find that our \oiii\ LF measurement at $z\sim 6.2$ is higher than the $z=3.24$ one \citep{khostovan15} by a factor of \textred{$\sim5$}, and the scenario of constant \oiii\ LF can be ruled out from $\chi^2$ statistics (\textred{$p$-value\,$<0.05$}).
This indicates a strong evolution of \oiii\ line strength and therefore possibly of ionization parameter towards higher redshift. 
\textred{Our \oiii\ LF measurement is also higher than the $z\sim8$ estimate in \citet{debarros19} by a factor of $\sim2$}.
\textred{From $\chi^2$ statistics}, these \oiii\ LF determinations at $z\sim6.2$ and $\sim$8 may be consistent with each other \textred{($p$-value\,$=0.3$; see also \citealt{matthee23} for most recent results).}
% because the small number statistics (this work) and limited accuracy of line luminosity measurement from broad-band SED modeling \citep{debarros19} could both result in considerable uncertainties.
% {\color{red} (If you want to include the following text, move it to Section 6.)}
% \st{Further JWST/NIRCam slitless spectroscopy surveys with F356W filter (e.g., ASPIRE and CEERS) and F444W filter (e.g., FRESCO) will provide more accurate constraints on the evolution of} \oiii\ \sout{LFs from $z\sim6$ to $z\sim8$.}

As shown in the top-right panel of Figure~\ref{fig:08_lf}, we also compare our \oiii\ LF with those predicted by the IllustrisTNG simulation at $z=5$ and 8 \citep{shen20} and the JAGUAR mock catalog \citep{williams18}.
These simulations/realizations generally underpredict the observed \oiii\,$\lambda$5007 LF by a factor of \textred{$\sim$10}.
Because our $\Phi_\mathrm{corr}(L)$ measurement of the \oiii\ LF in the bright bin ($10^{42.8}-10^{43.2}$\,\si{erg.s^{-1}}) is based on three sources (instead of only one) and the completeness correction factor is only \textred{$\sim1.4$}, we conclude that such a large excess of the \oiii\ LF compared to those in the simulation is real \textred{(see also recent simulations presented by \citealt{wilkins23})}.
% Although the reason of underprediction in simulations is beyond the scope of this study, we suspect that the ionization parameters of high-redshift luminous galaxies were likely underestimated, if low-redshift $\log{U} - Z/Z_\mathrm{\odot}$ relation was adopted (e.g., \citealt{carton17} relation matched in the JAGUAR mock catalog).
These simulations are typically tuned to match empirical distributions where possible, which in the case of ionization parameter ($\log{U}$), are best characterized at low redshifts (e.g. the $\log{U} - Z/Z_\mathrm{\odot}$ relation from \citealt{carton17}, in the case of the JAGUAR mock catalog). 
This is now known to underpredict rest-frame optical line fluxes such as \ha\ and \oiii\ at high redshifts \citep[e.g.,][]{debarros19,maseda19} as we find here, indicating the likely strong evolution that will be revealed by larger samples in upcoming JWST surveys.

\subsection{\ha\ Luminosity Function}
\label{ss:05c_half}

In the lower-left panel of Figure~\ref{fig:08_lf}, we first compare our $z\sim6.2$ \ha\ LF measurement (uncorrected for dust attenuation) with those at $z\sim0.4$  (\citealt{ly07}, \citealt{sobral13}; through 921\,nm narrow band) and $z\sim2.2$ (\citealt{geach08}, \citealt{hayes10}, \citealt{tadaki11}, \citealt{lee12}, \citealt{sobral13}), which are still accessible from the ground using the 2.12\,\micron\ narrow-band filter.
No direct measurement of \ha\ LF exists beyond $z\sim2.2$ before the launch of JWST.
We find that the $z\sim6.2$ \ha\ LF measured with this work is higher than those at $z\sim0.4$ but \textred{only slightly lower than} those at $z\sim2.2$, potentially suggesting weak or no evolution of \ha\ LF \textred{($p$-value\,$=0.36$ from $\chi^2$ statistics)} from the end of EoR to the peak of cosmic star-formation history ($z\sim2$, see \citealt{md14} for a review).
In the context of decreasing cosmic SFR density from $z\sim2$ to 6, the observed constancy of the \ha\ LFs between these two epochs possibly indicates, (\romannumeral1) a higher ionizing photon production efficiency at higher redshift (see discussion in Section~\ref{ss:04e_ha} and reference therein), which leads to a higher $L_\mathrm{H\alpha} / \mathrm{SFR}$ ratio in the EoR, and (\romannumeral 2) a decreasing obscured fraction of \ha\ emission, and therefore decreasing obscured fraction of cosmic SFR density from $z\sim2$ to 6 as shown in recent ALMA dust-continuum source surveys \citep[e.g.,][]{bouwens20,casey21,zavala21,sun22a}.

In the lower-right panel of Figure~\ref{fig:08_lf}, we show that our $z\sim 6.2$ \ha\ LF measurement is \textred{6--8} times higher than those predicted in the IllustrisTNG simulation ($z\sim5$; \citealt{shen20}) and JAGUAR mock catalog ($z\sim6.2$; \citealt{williams18}), similar to the \oiii\ LF.
As discussed in Section~\ref{ss:05b_o3lf}, this reflects our limited understanding of rest-frame optical emission line strength of $z\gtrsim 6$ galaxies before the launch of JWST.
This is likely caused by the underestimate of \textred{ionizing luminosities} for $z\gtrsim6$ galaxies with sub-solar metallicities \citep{williams18} like those in our sample.

In addition to this, if we simply translate the $z\sim6$ UV LF (e.g., \citealt{bouwens21a}) to \ha\ LF assuming the constant SFR conversions in \citet{ke12}, the volume density of \ha\ emitters would be underpredicted by $\sim 100$ times at \textred{$L_\mathrm{H\alpha}>10^{42.6}$\,\si{erg.s^{-1}}.}
This suggests that, (\romannumeral1) the dust attenuation is still substantial for UV-luminous galaxies even at $z > 6$ as is shown in recent ALMA studies \citep[e.g.,][]{schouws21,algera22,inami22}, despite a decreasing volume density of obscured SFR, and (\romannumeral2) an enhanced $\xi_\mathrm{ion}$ at high redshift when compared with the local Universe.
If we assume a typical $\log[\xi_\mathrm{ion}/(\mathrm{erg}^{-1}\,\mathrm{Hz})] = 25.5$ for galaxies at $z\sim6$ (see Section~\ref{ss:04f_xi}), the observed \ha\ LF can be well matched to the conversion from the UV LF with a color excess of $E(\mathrm{UV} - \mathrm{H\alpha}) = 0.9\pm0.5$.
This can be translated to an \textred{$E(B-V)=0.33\pm0.18$ and $A_V = 1.3\pm0.7$} assuming \citet{calzetti00} extinction law, and also \textred{$E(B-V)=0.11\pm0.06$ and $A_V = 0.30\pm0.17$} assuming the SMC extinction law of \citet{gordon03}, both adopting the canonical $E(B-V)$ ratio of 0.44 between stellar continuum and nebular lines \citep{calzetti00}.

\begin{figure}
\centering
\includegraphics[width=\linewidth]{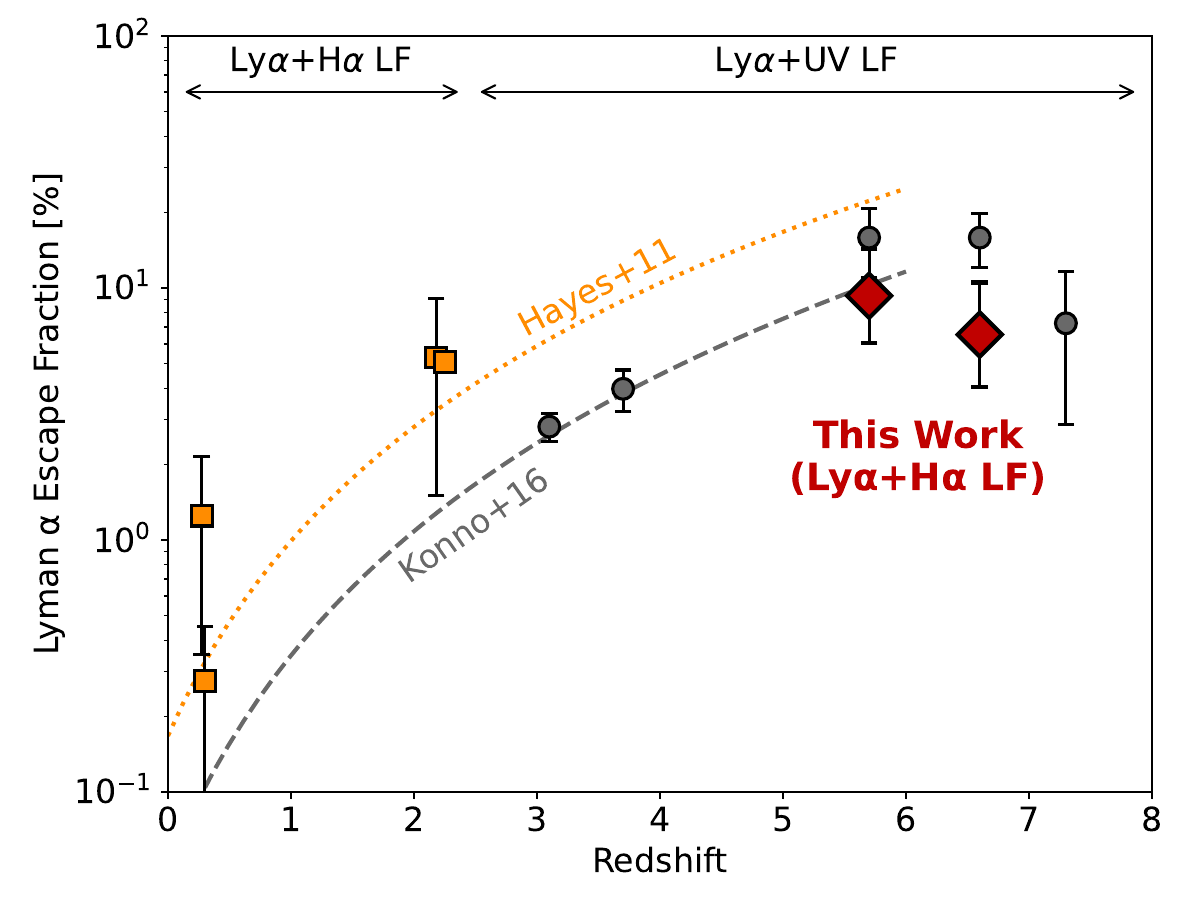}
\caption{Global escape fraction of \lya\ photons versus redshift. 
\lya\ escape fraction at $z=5.7$ and 6.6 based on \lya\ LF in \citet{konno18} and \ha\ LF measured in this work are shown as red diamonds (see Section~\ref{ss:05c_half}). 
$z\simeq0.2 - 2.2$ measurements using \lya\ and \ha\ LFs are shown in orange squares \citep{deharveng08,cowie10,hayes10b,hayes11,sobral17}, and the best-fit relation for the $z\simeq 0 - 6$ compilations in \citet{hayes11} is shown as the dotted orange line.
$z\simeq 3 - 8$ measurements using \lya\ and UV LFs compiled by \citet{konno16} are shown in gray circles, and the best-fit relation for their $z\simeq 0 - 6 $ compilation is shown as the dashed gray line.
}
\label{fig:09_lya}
\end{figure}

\subsection{Implication for the Global \lya\ Escape Fraction}
\label{ss:05d_lya}

As shown in the lower-right panel of Figure~\ref{fig:08_lf}, our observed \ha\ LF can match the observed \lya\ LF at $z=6.6$ \citep{konno18} assuming a \lya/\ha\ flux ratio of \textred{$0.57_{-0.22}^{+0.35}$}.
\textred{We note that this also assumes that both \lya\ and \ha\ emission is from the same population of star-forming galaxies.}
The best match to the \lya\ LF at $z=5.7$ \citep[also][]{konno18} would indicate a \lya/\ha\ flux ratio of \textred{$0.81_{-0.28}^{+0.44}$}.
Considering a theoretical \lya/\ha\ flux ratio of 8.7 from Case B recombination \textred{(see \citealt{hayes11} and \citealt{henry15})} and assuming no evolution of \ha\ LF from $z=5.7$ to 6.6, this suggests a global \lya\ escape fraction ($f_\mathrm{Ly\alpha}^\mathrm{esc} = f(\mathrm{Ly\alpha}) / f(\mathrm{H\alpha}) / 8.7$) of \textred{$6.5_{-2.4}^{+4.0}$\%} at $z\sim6.6$ and \textred{$9.3_{-3.2}^{+5.0}$\%} at $z\sim5.7$.
\textred{This is consistent with recent JWST measurements of $f_\mathrm{Ly\alpha}^\mathrm{esc}$ based on \lya-selected galaxies at $z\sim6$ \citep{ning23}.}
Note that these values are computed from the observed \ha\ LF, and the intrinsic \ha\ LF at $z\sim6.2$ could be even higher because of the dust extinction correction, which may lead to a potential overestimate of $f_\mathrm{Ly\alpha}^\mathrm{esc}$.

As shown in Figure~\ref{fig:09_lya}, both values of $f_\mathrm{Ly\alpha}^\mathrm{esc}$ are consistent with the global \lya\ escape fraction found at $z\sim2.2$ through the direct comparison of \lya/\ha\ LFs \citep[e.g.,][]{hayes10b,sobral17}, but only \textred{$\sim1/4$} of the early measurement of $f_\mathrm{Ly\alpha}^\mathrm{esc}$ at $z\sim 6$ in \citet[][$\sim30$\%]{hayes11} and \textred{$\sim1/2$} of that in \citet[][$\sim15$\%]{konno16}.
As discussed in \citet{konno16}, the difference in measured $f_\mathrm{Ly\alpha}^\mathrm{esc}$ between \citet{hayes11} and \citet{konno16} is mainly from the different lower limits of \lya\ and UV luminosities for the integration of their densities.  
In previous literature, the estimate of \lya\ escape fraction at $z>2.2$ relied on the conversion between UV and \ha\ SFR because direct \ha\ luminosity measurement was impossible from the ground. 
However, the \ha\ luminosity could be underestimated from SFR with the classical \citet{ke12} conversion because of an enhanced $\xi_\mathrm{ion}$ at higher redshift, leading to an overestimated \lya\ escape fraction (see also \citealt{matthee22}, \citealt{naidu22}). 
% {\color{red} (This sentence is a bit confusing.  If we assume the classical conversion factor, we would {\em underestimate} \ha\ and therefore overestimate \lya\ escape fraction.  Is this correct?)}

Our measurement further suggests no or weak evolution of \lya\ escape fraction at $z\simeq 2 - 6$, in contrast to previous estimates that generally predict a $f_\mathrm{Ly\alpha}^\mathrm{esc} \propto (1+z)^{2.6} \sim (1+z)^{2.8}$ evolution (see \citealt{hayes11}, \citealt{konno16} and a review by \citealt{ouchi20}).
Despite this, the general picture of declining $f_\mathrm{Ly\alpha}^\mathrm{esc}$ at $z>6$ when compared with those at $z=3.1$ and 5.7 \textred{\citep[e.g.,][]{konno16,konno18}} would still remain valid, which is consistent with the increasing neutral fraction of intergalactic medium at higher redshifts.
A more solid assessment would come in the future by obtaining \lya\ spectroscopy of the same targets, which could provide direct information on \lya/\ha\ line ratio.

\begin{figure}[!t]
\centering
\includegraphics[width=\linewidth]{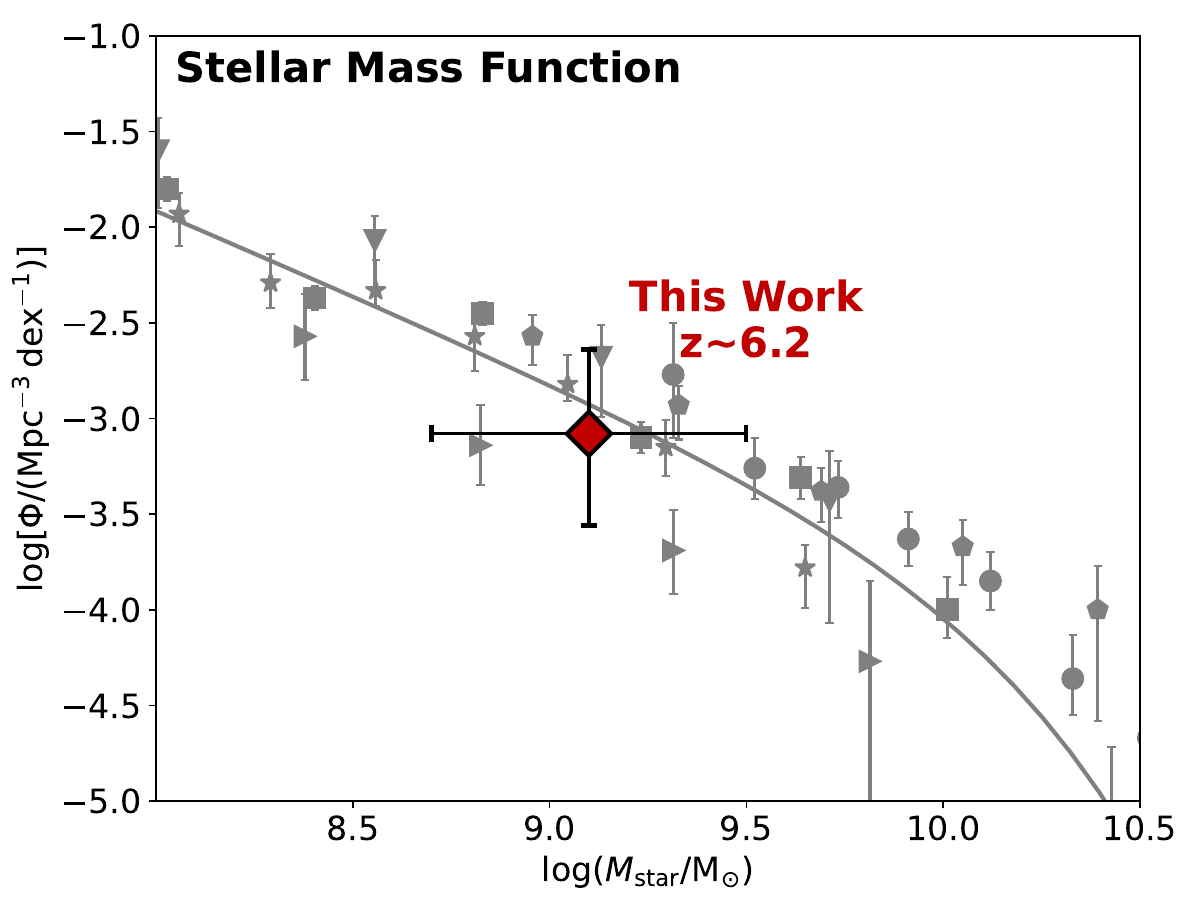}
\caption{Comparison with galaxy stellar mass function at $z\sim6$ (\citealt{duncan14}, pentagons; \citealt{grazian15}, circles; \citealt{song16}, rightward triangles; \citealt{bhatawdekar19}, downward triangles; \citealt{kikuchihara20}, stars; \citealt{stefanon21}, squares).
Compared with the best-fit Schechter function in \citet{stefanon22} as shown in the solid gray line, our sample recovers $88_{-57}^{+164}$\%\ of $z\sim6$ galaxies with stellar mass above $5\times10^8\,$\msun.
}
\label{fig:10_smf}
\end{figure}

\subsection{Biases \& Comparison with Stellar Mass Function}
\label{ss:05e_smf}

The selection of both \oiii\,$\lambda$5007 and \ha\ emitters could potentially bias our sample toward the high-metallicity end of emission-line galaxies at $z>6$, because metal-poor galaxies could have a lower \oiii/\ha\ flux ratio \citep[e.g.,][]{harikane18} and therefore the \oiii\ line could be fainter than the detection limit.
Meanwhile, the requirement of \ha\ detection can also bias our sample toward sources with higher instantaneous SFR and therefore younger stellar population, and sources with high \oiii\ luminosities (higher ionization parameters) but low \ha\ luminosities can be potentially missed.
In addition, the detection of line emitters requires the detection in direct imaging data, i.e., brighter than $\sim$25\,AB mag in either the F322W2 or F444W band. 
This means that for a source with comparable line fluxes, such as those in our sample, it would not make the selection if the stellar mass was $\sim 0.5$\,dex lower, leading to a selection bias against sources with large EWs \citep[e.g.,][]{maseda18}.

Bearing all of these potential biases in mind, we compare the volume densities of galaxies in our sample (using $1/V_\mathrm{max}$ method as in Section~\ref{ss:05a_method}) with those of $z\sim6$ galaxies with the similar stellar masses in Figure~\ref{fig:10_smf}.
The published samples include measurements from \citet{duncan14}, \citet{grazian15}, \citet{song16}, \citet{bhatawdekar19}, \citet{kikuchihara20} and \citet{stefanon21}, and all their stellar masses have been converted to those with \citet{chabrier03} IMF.
With a mean stellar mass of \textred{$1.4_{-0.5}^{+0.7}\times10^9$\,\msun}, the total volume density of galaxies in our sample is $10^{-3.1\pm0.5}$\,\si{Mpc^{-3}.dex^{-1}}.
Such a volume density is derived assuming a stellar-mass bin from \textred{$4\times10^8 \sim 4\times 10^9$\,\msun}. 
We also consider the error of stellar-mass measurements and small number statistics through MC simulation (similar to Section~\ref{ss:05a_method}).

The volume density of emission-line galaxies in our sample is \textred{$0.7_{-0.5}^{+1.2}$} times the volume density of $z\sim6$ galaxies with comparable stellar masses as measured in \citet{stefanon21} using their best-fit Schechter function.
By integrating the stellar-mass function in \citet{stefanon21}, we find that our sample recovers \textred{$66_{-44}^{+128}$\%} of the $z\simeq 6.0 - 6.6$ galaxies in the effective survey volume with stellar masses greater than $5\times10^{8}$\,\msun.
Such a high fraction of recovery suggests that our sample well represents \textred{a reasonably large subsample of $z\sim 6$ galaxies with  \mstar\,$\sim10^9$\,\msun}, and the emission-line galaxies are ubiquitous within such stellar mass and redshift ranges.

\section{Summary} \label{sec:06_con}

We present a sample of four \ha+\oiii\,$\lambda$5007 line emitters at $z=6.11-6.35$ discovered through JWST/NIRCam wide-field slitless spectroscopic observations obtained during the commissioning phase.
Located in the field of the flux calibrator P330-E, all four galaxies in our sample were spectroscopically confirmed with robust detections of \ha\ ($>3\sigma$) and \oiii\,$\lambda$5007 lines ($>5\sigma$), including the one at $z=6.11$ that has already been reported in \citet[\citetalias{sun22}]{sun22}.
\oii\,$\lambda$3727 and \oiii\,$\lambda$4959 lines were also tentatively identified for a few sources.
With the spectroscopic and photometric measurements obtained with JWST/NIRCam, we performed SED modeling, and analyzed their physical properties and volume densities. 
The main results are the following:

\begin{enumerate}

\item The median \hb$+$\oiii\ line EW of galaxies in our sample is \textred{$416\pm66$}\,\AA, consistent with those inferred previously at $z\simeq6.7-7.0$ from Spitzer/IRAC SED analysis \citep[e.g.,][]{endsley21b,endsley21a}. 
The median \ha\ line EW is \textred{$239\pm45$}\,\AA, smaller than those inferred previously at $z\simeq5.1-5.4$ in \citet[also from IRAC SED analysis]{rasappu16}.
All \oiii\ and \ha\ EWs of these $z>6$ galaxies are much larger than those in the local Universe with similar stellar masses.

\item Sources in our sample likely occupy the same parameter space as that of $z\sim2$ star-forming galaxies in the \oiii/\hb-\nii/\ha\ BPT diagram, but are located above the star-forming sequence of local galaxies in the diagram.
This could be explained by their elevated N/O abundance at a given O/H ratio, and/or higher ionization parameters, like those seen with $z\sim2$ galaxies.

\item Through the strong-line ratios, we show that these $z>6$ galaxies have been enriched to moderate metallicities (\textred{$\sim0.4$\,\si{Z_\odot}}), comparable to those of galaxies \textred{at $z=2-3$} with similar stellar masses.
Combined with the direct metallicity measurements of low-mass galaxies obtained from the JWST/NIRSpec ERO data, we find a steep slope of the mass-metallicity relation in the EoR.
This may indicate a rapid metallicity enrichment history in certain massive ($M_\mathrm{star}\gtrsim10^9$\,\msun) galaxies at $z>6$.

\item The median \ha\ EW of galaxies in our sample is higher than that of \textred{star-forming} galaxies at $z\lesssim 2$, but is consistent with that at $z\simeq4-5$ inferred previously.
This may indicate a flattening of the redshift evolution of \ha\ EWs towards the EoR. 
% than certain previous estimates.

\item The galaxies in our sample have a median ionizing photon production efficiency of $\log[\xi_\mathrm{ion}/(\mathrm{erg}^{-1}\,\mathrm{Hz})] = 25.4\pm0.2$.
Even with some uncertainty associated with dust attenuation correction, this value appears consistent with the redshift evolution trend in previous studies where $d\log(\xi_\mathrm{ion})/dz = 0.10\pm 0.02$.
A higher $\xi_\mathrm{ion}$ at higher redshift can lead to a higher \ha-based SFR estimate when compared to that derived from UV, if the canonical conversion factors are used \citep[e.g.,][]{ke12}.

\item We derived the \oiii\,$\lambda$5007 and \ha\ line luminosity functions using the $1/V_\mathrm{max}$ method.
This is the first time that these two line LFs can be directly measured in the EoR.
The $z\sim6.2$ \oiii\ LF we measured is higher than those measured previously at $z=3.24$ \citep{khostovan15}, \textred{but likely consistent with that} inferred at $z\sim8$ \citep{debarros19}.  It is also higher than those in the IllustrisTNG simulation \citep{shen20} and JAGUAR mock catalog \citep{williams18} at comparable redshift by \textred{$\sim10$} times.

\item Our measurements suggest \textred{weak or no} evolution with the \ha\ line LF from $z\sim2$ to 6.  The measured \ha\ LF at $z\sim6$ exceeds the predictions from certain previous simulations/realizations by a factor of \textred{6--8}.
A simple conversion from the $z\sim6$ UV LF assuming SFR(\ha)\,$=$\,SFR(UV) would underpredict \ha\ LF (at \textred{$L_\mathrm{H\alpha}>10^{42.6}$\,\si{erg.s.^{-1}}}) by a factor of $\sim100$.
This suggests an enhanced ionizing photon production efficiency and a substantial dust attenuation for UV-luminous galaxies at $z\sim6$.

\item By directly comparing the \lya\ and \ha\ LF at $z\sim6$, we find a global \lya\ escape fraction of \textred{$9.3_{-3.2}^{+5.0}$\%} at $z=5.7$ and \textred{$6.5_{-2.4}^{+4.0}$\%} at $z=6.6$.
In contrast to previous studies that used UV-based SFR to infer \ha\ and therefore intrinsic \lya\ luminosity, our study suggests no or weak evolution of the \lya\ escape fraction at $z\simeq 2- 6$.
Despite this, the general picture of declining $f_\mathrm{Ly\alpha}^\mathrm{esc}$ at $z>6$ should still remain valid.

\item Despite some potential biases inherent in our selection of \oiii$+$\ha\ emitters, the four galaxies in our sample have already contributed to a volume density of \textred{$10^{-3.1\pm0.5}$\,\si{Mpc^{-3}.dex^{-1}}}.
This means that our sample recovers \textred{$66_{-44}^{+128}$\%} of the $z= 6.0 - 6.6$ galaxies within the effective survey volume with $M_\mathrm{star} > 5\times 10^{8}$\,\msun. 
Such a high recovery fraction indicates a low selection bias of our sample as well as the ubiquity of emission-line galaxies within such stellar-mass and redshift ranges.

\end{enumerate}

% Also note that the acknowledgment environment does not support long amounts of text. If you have a lot of people and institutions to acknowledge, do not use this command. Instead, create a new 
\section*{Acknowledgments}
% \begin{acknowledgments}
\textred{We thank the anonymous referee for their helpful comments.}
F.S.\ thanks Fuyan Bian, Zheng Cai, Daniel Ceverino and Daniel P.\ Stark for helpful discussions.
F.S., E.E., M.R., D.J.E., D.K., C.C.W.\ and C.N.A.W.\ acknowledge funding from JWST/NIRCam contract to the University of Arizona, NAS5-02105.
A.J.B.\ acknowledges funding from the “FirstGalaxies” Advanced Grant from the European Research Council (ERC) under the European Union’s Horizon 2020 research and innovation program (Grant agreement No. 789056).

This work is based on observations made with the NASA/ESA/CSA James Webb Space Telescope. The data were obtained from the Mikulski Archive for Space Telescopes at the Space Telescope Science Institute, which is operated by the Association of Universities for Research in Astronomy, Inc., under NASA contract NAS 5-03127 for JWST. These observations are associated with program \#1076.
% \end{acknowledgments}
{This paper is based upon High Performance Computing (HPC) resources supported by the University of Arizona TRIF, UITS, and Research, Innovation, and Impact (RII) and maintained by the UArizona Research Technologies department.}
{All of the data presented in this paper were obtained from the Mikulski Archive for Space Telescopes (MAST) at the Space Telescope Science Institute. The specific observations analyzed can be accessed via \dataset[10.17909/f8p1-e696]{https://doi.org/10.17909/f8p1-e696}. }


% \vspace{0mm}
\facilities{JWST (NIRCam)}


\software{\textsc{astropy} \citep{2018AJ....156..123A},  \textsc{Source Extractor} \citep{1996A&AS..117..393B}, \textsc{photutils} \citep{photutils}, \textsc{cigale} \citep{boquien19}
          }



\appendix

\section{Completeness and Effective Survey Volume}
\label{apd:01}

\renewcommand{\thefigure}{A\arabic{figure}}
\renewcommand{\theHfigure}{A\arabic{figure}}
\setcounter{figure}{0}

\begin{figure}
\centering
\includegraphics[width=0.325\linewidth]{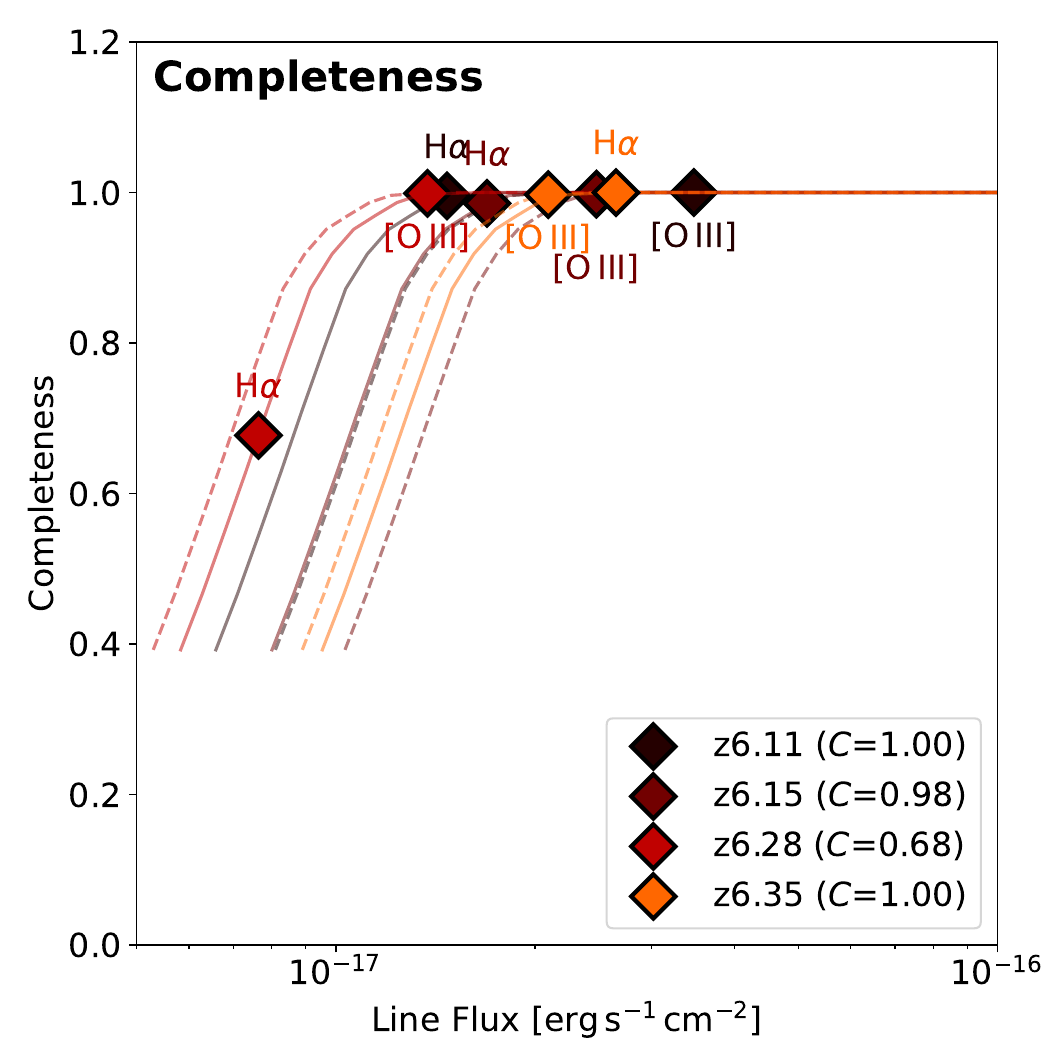}
\includegraphics[width=0.325\linewidth]{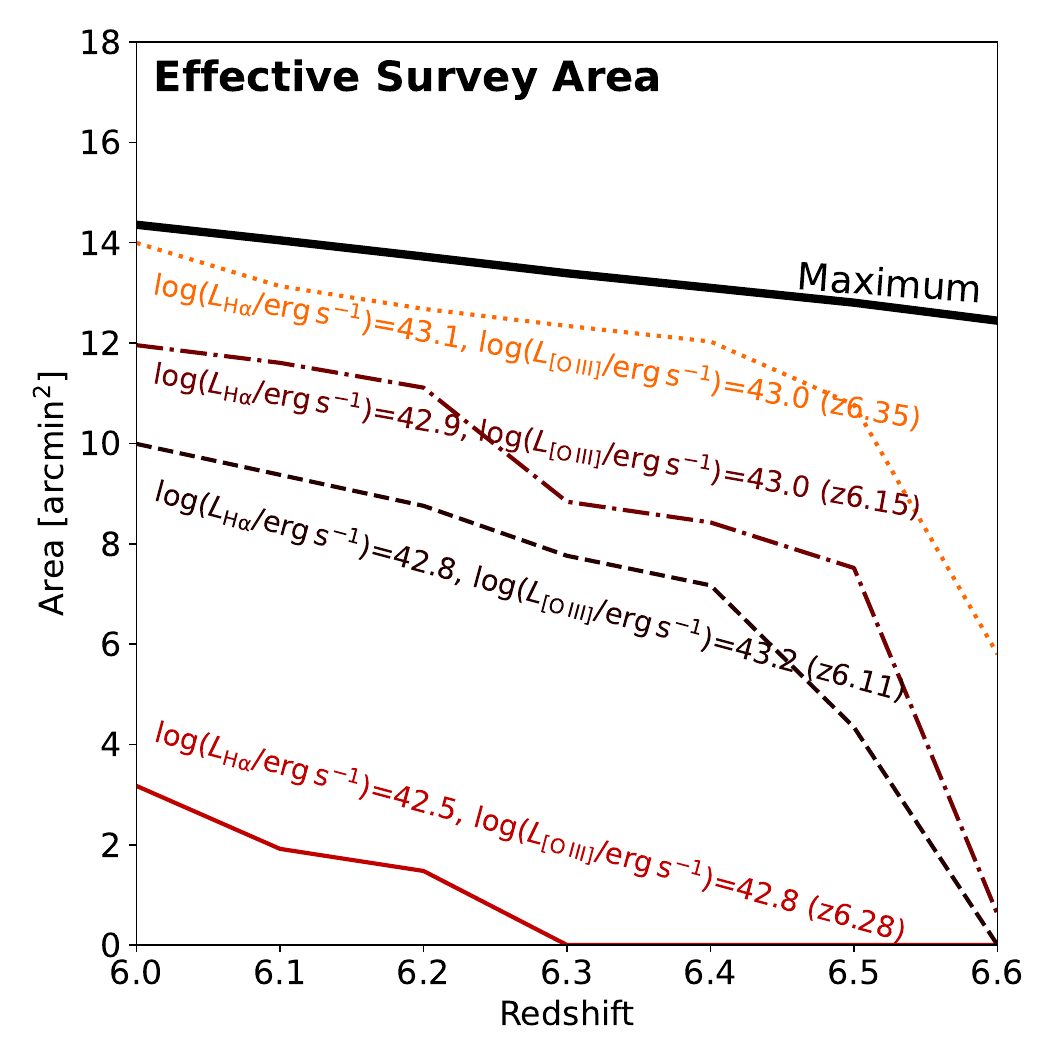}
\includegraphics[width=0.325\linewidth]{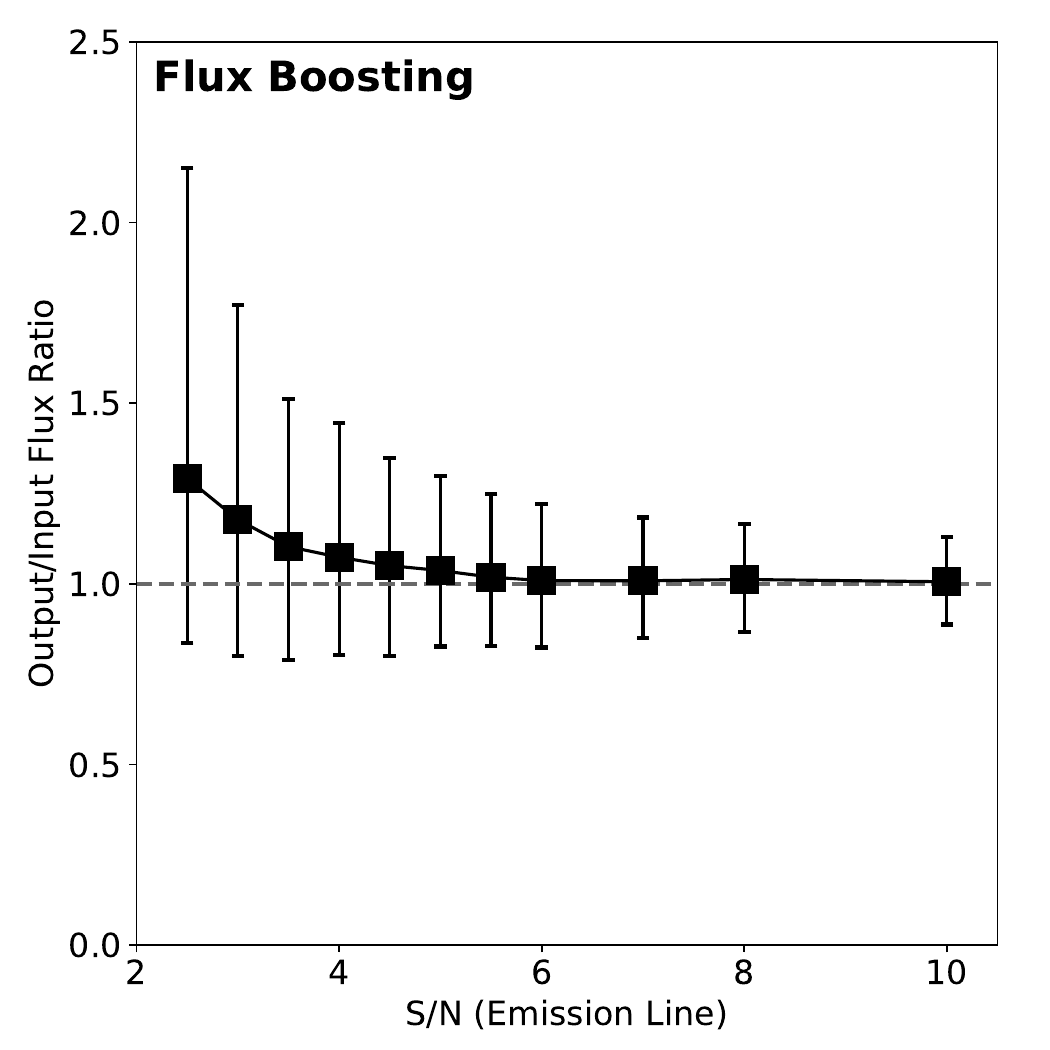}
\caption{Left: Completeness of line emission detection. 
For each source at $z=6.11-6.35$, we show the simulated completeness as a function of input line flux according to the depth of their observation setup. 
Completeness functions for \ha\ and \oiii\ lines are shown as solid and dashed curves, respectively. % (see Appendix~\ref{apd:01}).
The synthesized completeness of line detections for each source is shown in the figure legend.
Middle: Effective survey area as a function of redshift. 
The maximum survey area is shown as the solid black line, and effective survey area for four sources in our sample are shown as colored curves computed from the measured line luminosities.
Right: Flux boosting effect (output/input line flux ratio) as a function of input line S/N modeled in our Monte Carlo simulation.
The error bars represent 16th and 84th percentiles of the output flux ratios for each realization.
}
\label{fig:apd_01}
\end{figure}

\subsection{Completeness}
\label{apd:a1}

To compute the completeness of all \oiii\,$\lambda$5007 and \ha\ line detections, we first select faint continuum sources (F444W\,$>24$\,AB\,mag) for which the grism observations yielded similar spectral coverages as those of $z>6$ line emitters.
These continuum sources were observed with the same numbers of integrations with each module/grism combination (AR, BR, AC and BC) at the wavelength of interest (i.e., \oiii\,$\lambda$5007 and \ha\ wavelengths at $z\sim6.2$) as those of each line emitter, respectively.
However, these F444W sources have no continuum or line emission detected in the grism data.

We then extract and coadd their 2D spectral image, and inject mock line emissions randomly at the same wavelength range of \oiii\,$\lambda$5007 and \ha\ lines at $z=5.9-6.4$.
No mock line emissions are injected at $z>6.4$ \oiii/\ha\ wavelengths. 
This is because the observed \ha\ wavelengths at these redshifts are too close to the red end of the F444W filter, and therefore the number of coadded integrations can decrease significantly in the 2D spectral images of these faint continuum sources, resulting in jumps of RMS noise.
The injected line emissions are represented as 2D Gaussian profiles with FWHM\,$=5\sim6$\,pixel in both directions, similar to those of the sources in our sample.
The line fluxes and errors are measured using the same method as we applied for the real line detections.
The injected line emissions are at various strengths that yield detections at significance levels, ranging from $2\sigma$--$100\sigma$.

Based on the median output line flux errors, we are able to derive the fraction of $\geq 3\sigma$ detections for each of the \oiii/\ha\ line fluxes in our sample, which is the completeness of each line detection. 
In reality, we find that the fraction of $\geq 3\sigma$ detections is well correlated with the S/N of input line, which is defined as the input line flux divided by the median output line flux error.
For injected line emissions with input S/N of 5, the completeness of $3\sigma$ detection is 97\%. 
However, this decreases to 58\% for injected emission lines with a input S/N of 3. % note eddington bias, 50% -> 58%
Such a completeness function of input line S/N is used to compute the completeness of line detections in our sample.
The left panel of Figure~\ref{fig:apd_01} shows the completeness of all line detections as a function of the line fluxes.
Except for the \ha\ line of P330E-z6.28 that has a completeness of $\sim$60\%, the completeness of all of the other lines are close to 100\%. 
The synthesized completeness is computed as the product of \oiii\ and \ha\ line completeness, which is given in the lower-right corner of the same panel.

\subsection{Survey Volume}
\label{apd:a2}
The maximum survey volume of each source in our sample is computed from the RMS map of F322W2 and F444W grism images. 
We first measure the RMS noise (unit: DN/s) of individual grism integrations taken with each of the four module/grism combinations in both F322W2 and F444W band, respectively.
We then constructed the mosaicked RMS maps for grism images based on the measured noises and the pointing information of the telescope.
We are then able to convert these to the maps of $3\sigma$ detection threshold for \oiii\ and \ha\ line luminosities at $z=6.0\sim6.6$ using (\romannumeral1) the sensitivity curves of four module/grism combinations in F322W2/F444W filter, and (\romannumeral2) the relation between the RMS noise of 2D spectral image and the median line flux error obtained from the simulations described above.
For each source in our sample, we compute the overlapping area from $z=6.0$ to $z=6.6$ at a step of $\Delta z = 0.1$, where both their \oiii\ and \ha\ line luminosities are above the detection threshold.

The maximum survey areas for all four sources in our sample as functions of redshift are shown in the middle panel of Figure~\ref{fig:apd_01}.
With the lowest \ha\ and \oiii\ luminosities, source P330E-z6.28 can only be detected in the deepest 1--2\,\si{arcmin^2} area of the survey that contains the flux calibrator, consistent with what we observed.
The maximum survey volume is then integrated from the maximum survey area across $z=6.0-6.6$.
We also run MC simulations to quantify the errors of maximum survey volume propagated from the uncertainty of line flux, which is small ($\sim$\,5\%) for luminous emitters like P330E-z6.35 but large ($\sim$100\%) for faint emitters like P330E-z6.28. 
We also consider the reduction of maximum survey volume caused by the existence of bright continuum sources in the field.
We conclude that the bright continuum contamination will only result in a small reduction ($\lesssim 5$\%) of maximum survey volume from visual inspection of the extracted spectra of faint continuum sources (F444W\,$\gtrsim24$\,AB mag).
This is because both R and C grism data are available and the field is sparse of bright stars (galactic latitude $b=+42$\arcdeg).
Such a small reduction is not worthy of correction when compared with the large uncertainty from small number statistics.

% From the pointing information of the telescope and the RMS noises of individual F322W2 and F444W grism integrations taken with 
% We also evaluate the flux boosting effect and the Eddington bias by comparing injected and output line flux ratios, and 
% We also compute the maximum survey volume of each source from the RMS map of F322W2 and F444W grism images, in which the \oiii\,$\lambda$5007 and \ha\ lines with the same luminosities as those of the real source can be detected at $\geq 3\sigma$.
% In general, the maximum survey area decreases from $z=6.0$ to 6.6 because of the increasing RMS noise toward the red end of F444W filter. 

\subsection{Flux Boosting and Eddington Bias}
\label{apd:a3}
The flux measurements of low-significance emission lines often suffer from the flux-boosting effect because the Gaussian line-profile fitting routine always searches for the peak of the signal and returns a positive flux.
If uncorrected, this will result in an overestimate of the volume density of luminous sources, also known as the Eddington bias \citep{eddington13}.
The flux-boosting effect is quantified by comparing the injected and output line fluxes at input line S/N from 2 to 100 (Appendix~\ref{apd:a1}).
For each input line S/N, we study the 16-50-84th percentiles of the output/input line flux ratios, which is shown as the right panel of Figure~\ref{fig:apd_01}.
Such a flux-boosting effect is corrected for our line luminosity and LF measurements, and the line luminosity could be boosted by 18\%\ for a S/N\,$=$\,3 detection but only 5\%\ at S/N\,$=$5.
This is also the reason why 58\% of lines with input S/N\,$=$\,3 are actually detected at S/N\,$\geq$\,3, instead of 50\%, as described in the Appendix~\ref{apd:a1}.


\bibliography{00_main}{}
% \bibliography{99_ref,High-z_Galaxies_ads,NIRCam_Grism_ads}{}
\bibliographystyle{aasjournal}



\end{document}